\documentclass[
twocolumn,
preprintnumbers,superscriptaddress,showpacs,supberscriptaddress,
nofootinbib,longbibliography]{revtex4-2}

\usepackage[
bookmarks=true,colorlinks,linkcolor=blue,urlcolor=cyan,citecolor=red]{hyperref}

\usepackage{graphicx}
\usepackage{latexsym}
\usepackage{amsfonts}
\usepackage{amssymb}
\usepackage{amsmath}
\usepackage{bm}
\usepackage{mathrsfs}
\usepackage{slashed}
\usepackage{ascmac}
\usepackage{comment}

\usepackage{dcolumn}

\allowdisplaybreaks[3] 

\usepackage[utf8]{inputenc} 

\newcommand{\cout}[1]{ \if 0 {#1} \fi }
\newcommand{\nn}{\nonumber}
\renewcommand{\=}{&=&}
\newcommand{\nnb}{\nonumber \\}
\newcommand{\pd}{\partial}

\renewcommand{\r}{\right}

\renewcommand{\a}{\alpha}
\renewcommand{\b}{\beta}

\newcommand{\m}{\mu}
\newcommand{\n}{\nu}
\renewcommand{\r}{\rho}
\newcommand{\s}{\sigma}

\newcommand{\gam}{ \gamma }

\newcommand{\bk}{{\bm{k}}}

\newcommand{\bv}{{\bm{v}}}

\newcommand{\bB}{{\bm B}}

\newcommand{\bu}{{\bm u}}

\newcommand{\order}{{\mathcal O}}

\newcommand{\V}{ {\mathcal V} }

\newcommand{\para}{ \parallel}

\newcommand{\zero}{{(0)}}
\newcommand{\one}{{(1)}}

\usepackage{color} 
\usepackage[normalem]{ulem}  
\newcommand{\com}[1]{[{\color[rgb]{0,0,1}{#1}}]}

\newcommand{\cgd}[1]{{\color[rgb]{1,0,0}{#1}}}

\newcommand{\pp}[1]{{\color[rgb]{0.7,0.3,0.9}{#1}}}

\begin{document}

\title{Analytic solutions for the linearized first-order magnetohydrodynamics and implications for causality and stability }

\author{Zhe Fang}
\email{12345071@zju.edu.cn}
\affiliation{Zhejiang Institute of Modern Physics, Department of Physics, Zhejiang University, Hangzhou, Zhejiang 310027, China}

\author{Koichi Hattori}
\email{koichi.hattori@zju.edu.cn}
\affiliation{Zhejiang Institute of Modern Physics, Department of Physics, Zhejiang University, Hangzhou, Zhejiang 310027, China}
\affiliation{Research Center for Nuclear Physics, Osaka University, 
10-1 Mihogaoka, Ibaraki, Osaka 567-0047, Japan}

\author{Jin Hu}
\email{hu-j23@fzu.edu.cn}
\affiliation{Department of Physics, Fuzhou University, Fujian 350116, China}

\begin{abstract}
{
We address the linear-mode analysis performed near an equilibrium configuration in the fluid rest frame {with a dynamical magnetic field perturbed on a constant configuration.}
We develop a simple and general algorithm for an analytic solution search that works on an order-by-order basis in the derivative expansion. 
This method can be applied to general sets of hydrodynamic equations. 
Applying our method to the first-order relativistic magnetohydrodynamics, 
we demonstrate that the method finds a complete set of solutions. 
}
We obtain two sets of analytic solutions for the four and two coupled modes with 
seven dissipative transport coefficients. 
The former set has been missing in the literature for a long time {due to the difficulties originating from coupled degrees of freedom and strong anisotropy provided by a magnetic field}. 
{The newly developed method resolves these difficulties}. 
We also find that the small-momentum expansions of the solutions break down when the momentum direction is nearly perpendicular to an equilibrium magnetic field due to the presence of another small quantity, that is, a trigonometric function representing the anisotropy. We elaborate on the angle dependence of the solutions and provide alternative series representations that work near the right angle. 
{
This identifies the origin of a discrepancy found in recent works. 
}
Finally, we discuss the issues of causality and stability based on our analytic solutions and recent developments in the literature. 
\end{abstract}

\maketitle

\section{Introduction}\label{s1}

Relativistic magnetohydrodynamics (MHD) has been developing as a framework to describe various systems ranging from the femtoscale droplets realized in relativistic heavy-ion collisions at RHIC and LHC \cite{Inghirami:2016iru, Inghirami:2019mkc, Nakamura:2022idq, Nakamura:2022wqr, Nakamura:2022ssn} to the cosmological/astronomical scales. 
The latter includes the accretion flows and jet formation near black holes (BH) \cite{McKinney:2008ev, McKinney:2012vh, Bromberg:2015wra, Davis:2020wea}, supernova explosions \cite{Shibata:2006hr,Matsumoto:2020rbz, Matsumoto:2022hzg}, and the binary mergers of neutron stars (NSs) \cite{Duez:2005cj, Shibata:2021bbj, Shibata:2021xmo, Kiuchi:2023obe} 
and of NS-BH \cite{Etienne:2011ea, Hayashi:2021oxy} (see also references therein). 
More recent developments include observations of the magnetic inverse cascade with coupled dynamics of the magnetic helicity (and the fermion chirality) \cite{Boyarsky:2011uy, Rogachevskii:2017uyc, Brandenburg:2017rcb, Schober:2017cdw, Masada:2018swb, DelZanna:2018dyb, Matsumoto:2022lyb, Schober:2021iws, Brandenburg:2023aco, Brandenburg:2023rrd, Schober:2023zxl} (see Refs.~\cite{brandenburg2023chirality, Kamada:2022nyt, Hattori:2023egw} for reviews).

In recent years, relativistic MHD was reformulated based on the conservation of the magnetic flux \cite{Grozdanov:2016tdf, Hattori:2017usa, Armas:2018atq, Hongo:2020qpv} (see Ref.~\cite{Hattori:2022hyo} for a review). 
On the other hand, the conventional formulation is based on a coupled system of the Maxwell equation and the energy-momentum conservation law, where both equations have the source terms stemming from the electric current and the Joule heat and Lorentz force, respectively. This implies the presence of nonconserved charges, i.e., gapped modes, involved in the conventional formulation. 
In fact, electric fields are damped out or screened when systems approach equilibrium states. 
The new formulation canonically follows the spirit of hydrodynamics, that is, conservation laws associated with symmetries. 
The magnetic flux conservation is identified as a consequence of the generalized concept of global symmetries called the magnetic one-form symmetry \cite{Gaiotto:2014kfa}.

{
In this paper, we focus on solving the set of MHD equations {within the linear-mode analysis near an equilibrium state} and develop an analytic algorithm for the solution search on an order-by-order basis in the derivative expansion. 
We emphasize that our method works as a general algorithm for the solution search, and can be applied to general sets of equations based on derivative expansions. 
It is useful to obtain analytic solutions for hydrodynamics since the transport coefficients and the equations of state are often not (precisely) known in individual systems. 
Our method paves a new avenue toward analyses of causality and stability of MHD, especially, with extensions beyond the conventional first-order derivative expansion. 
In other words, without a general method for overcoming difficulties in the solutions search, stability and causality analyses may not be achieved.

We demonstrate our method by applying it to the first-order MHD of which the complete set of solutions has not been obtained in the literature. 
The difficulties in previous works stem from additional degrees of freedom and strong anisotropy provided by a magnetic field. 
The analyses in the literature are limited to two particular angles parallel and perpendicular to a magnetic field, or otherwise an isotropic limit. 
Moreover, there has been disagreement on the solution in the perpendicular direction. 
We resolve this issue with our complete set of analytic solutions and identify the origin of disagreement with the inherent anisotropy in MHD. 
It turns out that the disagreement signals a breakdown of native momentum expansion in a boarder range of the angle as well as at the right angle (see below). 
The anisotropy is a key feature of MHD that should not be ignored toward further developments.

More specific backgrounds are described as follows. }
We linearize the MHD equations with respect to perturbative disturbances applied to an equilibrium state and obtain the dispersion relations for the eigenmodes. 
This analysis, called the linear-mode analysis, has been conventionally used to discuss causality and stability of hydrodynamic theories \cite{Hiscock:1983zz,Hiscock:1985zz,Hiscock:1987zz} (see below for recent progress on causality and stability analyses). 
While the linear-mode analysis was applied to the relativistic MHD in the recent literature \cite{Grozdanov:2016tdf, Hernandez:2017mch,Biswas:2020rps, Armas:2022wvb}, a complete set of solutions is still missing. 
The difficulty simply arises from the fact that one needs to diagonalize a large matrix in the absence of a spatial rotational symmetry broken by a magnetic field.

The complete set of solutions consists of six gapless modes, which are known as a pair of the Alfv\'{e}n waves and two pairs of the magnetosonic waves. 
We fully include the first-order derivative corrections that consist of three bulk viscosities, two shear viscosities, and two electric resistivities. 
We focus on the Landau frame while a straightforward extension can be carried out for a general frame choice and general matching conditions of hydrodynamic variables  (see, e.g., Ref.~\cite{Hattori:2022hyo} for discussions about these choices in MHD and Ref.~\cite{Armas:2022wvb} for a recent linear-mode analysis without a specific choice). 
In the presence of the anisotropic corrections, the solutions for the magnetosonic modes had been known only in the two particular limits where the momentum is oriented parallel or perpendicular to a background magnetic field \cite{Grozdanov:2016tdf, Hernandez:2017mch, Armas:2022wvb}.

We further discuss the convergence of the small-momentum expansion by inspecting the obtained solutions and 
find that the higher-order terms in the small-momentum expansion diverge when the momentum is taken nearly perpendicular to a magnetic field. 
This issue is caused by the presence of another small quantity, that is, a trigonometric function representing the angle dependence. 
We find this issue both in the Alfv\'{e}n and magnetosonic modes. 
It can be a general issue in anisotropic systems. 
We identify the correct result obtained from the original equations before the small-momentum expansion is performed, giving a different result than that in the literature \cite{Grozdanov:2016tdf, Armas:2022wvb}. 
We provide an alternative series representation that correctly captures the anisotropy near the right angle.

Finally, we discuss the causality and stability of relativistic MHD. 
We show that the phase velocities, which are called the Alfv\'{e}n velocities and the fast and slow magnetosonic velocities, are always smaller than the speed of light. 
This means that the linear waves are causal {\it within} the ideal MHD. 
We also show that the first-order corrections, obtained within the fluid rest frame, always 
provide damping effects on the linear waves as long as the transport coefficients satisfy the inequalities required by the second law of thermodynamics. 
This is expected since MHD only contains dissipative transport coefficients.\footnote{In the strict hydrodynamic limit, MHD does not contain the Hall terms due to the absence of net electric charge density.} 
The absence of growing modes indicates stability of an equilibrium state {\it within the fluid rest frame} where the linear-mode analysis is performed. 
However, those conditions are not sufficient to guarantee the causality and stability of dissipative hydrodynamics in general Lorentz frames. 
It has been widely known that diffusive modes are acausal, leading to developments of the Israel-Steward theory \cite{Israel:1976tn, Israel:1978up, Israel:1979wp} (see, e.g., Ref.~\cite{Denicol:2014loa} for a review), 
and observers in different Lorentz frames could see (unphysical) unstable modes (see, e.g., Ref.~\cite{Denicol:2008ha}).

Recent developments deepened our understanding of the issues of causality and stability. 
A stronger necessary condition for a stable dispersion relation was obtained from complex analysis of a retarded propagator in general causal theories \cite{Heller:2022ejw}. 
Then, it was explicitly shown that, 
unless this necessary condition  
is satisfied, one finds a boost velocity that transforms a stable mode in the fluid rest frame to an unstable mode in the boosted frame 
\cite{Gavassino:2021owo, Gavassino:2023myj}. 
We briefly discuss a generalization of this crucial condition to anisotropic systems including MHD. 
The reader is referred to recent related works  \cite{Bemfica:2017wps, Bemfica:2019knx, Bemfica:2020zjp, Bemfica:2020xym} \cite{Kovtun:2019hdm, Hoult:2020eho} \cite{Wang:2023csj} that may be classified in terms of employed criteria for the causality and stability; with or without specific choices of a flow vector and hydrodynamic variables; and with or without the linearization. 
One is demanded to perform more analyses when including more conserved charges such as a vector charge \cite{Brito:2020nou}, magnetic field \cite{Biswas:2020rps, Armas:2022wvb,Cordeiro:2023ljz}, axial charge \cite{Speranza:2021bxf, Abboud:2023hos}, and/or spin \cite{Daher:2022wzf, Sarwar:2022yzs, Weickgenannt:2023btk, Xie:2023gbo}.

This paper is organized as follows. 
We first recapitulate the recent formulation of relativistic MHD in Sec.~\ref{s2} 
and show a set of linearized equations in Sec.~\ref{s3}. 
In Sec.~\ref{s4}, we introduce our method for the solution search. We elaborate on the convergence/breakdown of the small-momentum expansion. 
In Sec.~\ref{sec:s-and-c}, we discuss the causality and stability of MHD, which is supported by the Appendixes. 
Finally, we conclude this paper in Sec.~\ref{sec:conclusion}.  
Throughout the paper, we use the mostly plus metric convention ${\eta ^{\mu \nu }} = \text{diag}( - 1,1,1,1)$ and the completely antisymmetric tensor with the convention $\epsilon^{0123} =  + 1$. Then, the fluid velocity ${u^\mu }$ is normalized 
as ${u^\mu }{u_\mu } = - 1$. 
We define the projection operator ${\Delta ^{\mu \nu }} = {\eta ^{\mu \nu }} + {u^\mu }{u^\nu }$ such that $u_\mu \Delta^{\mu\nu}=0 $. 
To specify the direction of a magnetic field, we introduce a spatial unit vector ${b^\mu } $ such that $b^\mu b_\mu = 1$ and $u_\mu b^\mu =0$ and accordingly another projection operator ${\Xi ^{\mu \nu }} = {\Delta ^{\mu \nu }} - {b^\mu }{b^\nu }$ 
such that $b_\mu \Xi ^{\mu \nu }=0= u_\mu \Xi ^{\mu \nu }$.

\section{First-order MHD from the magnetic-flux conservataion}\label{s2}

We recapitulate the formulation of relativistic MHD with the magnetic-flux conservation \cite{Grozdanov:2016tdf, Hattori:2017usa, Armas:2018atq, Hongo:2020qpv} (see Ref.~\cite{Hattori:2022hyo} for a review). 
While the magnetic flux is conserved in the absence of 
a magnetic monopole, the electric flux can terminate 
at electric charges, implying that an electric field is 
not a conserved quantity. 
The conservation laws for the energy-momentum tensor 
$ \Theta ^{\mu \nu }$ and the (dual) electromagnetic field strength tensor ${\tilde F}^{\mu \nu } = \frac{1}{2}\epsilon^{\mu \nu\rho\sigma}F_{\rho\sigma} $ read 
\begin{equation}
{\partial _\mu }{\Theta ^{\mu \nu }} = 0 , \quad  
{\partial _\mu }{{\tilde F}^{\mu \nu }} = 0 . 
\label{con4}
\end{equation} 
Realizing the second equation as a conservation law 
for the magnetic flux led to a renewed formulation of magnetohydrodynamics along with the symmetry guideline, 
but with a generalized notion called the one-form symmetry.


The temporal components of the conserved currents 
provide the corresponding conserved charges 
\begin{eqnarray}
\label{eq:charges}
e =  u_\mu u_\nu {\Theta ^{\mu \nu }} , \quad  
B^\mu  = -  {\tilde F}^{\mu \nu } u_\nu . 
\end{eqnarray} 
We postulate that 
these quantities satisfy the first law of thermodynamics 
\begin{subequations}
\label{con5}
\begin{eqnarray} 
Ts \= e + {p }   - {H_\mu }{B^\mu } , \\
Tds \= de   - {H_\mu }d{B^\mu } ,
\\
TDs \= De   - {H_\mu }D{B^\mu },
\end{eqnarray}
\end{subequations}
where we defined the temporal derivative $D = u^\mu\partial_\mu$. 
{
We introduced pressure $p$ 
and the Lagrange multipliers, i.e., 
the (inverse) temperature $T$ 
and the magnetic field $H_\mu$ for the conserved energy density $e$ and the magnetic flux $B^\mu$, respectively. 
}
The translational symmetry of the system 
guarantees the conservation of 
the {\it total} energy density $e $ that contains 
not only matter contributions but also electromagnetic energy. 
The corresponding pressure $p $ should also be the total quantity.

To organize a closed system of equations, 
one needs to obtain the constitutive equations that express the spatial components of the conserved currents by the conserved charges. 
Based on the derivative expansion, 
the constitutive equations can be written down as 
\begin{subequations}
\label{con6}
\begin{eqnarray}  
{\Theta ^{\mu \nu }} \= e{u^\mu }{u^\nu } + {p_\parallel }{b^\mu }{b^\nu } + {p_ \bot }{\Xi ^{\mu \nu }} 
+ \Theta _{(1)}^{\mu \nu }, \\ 
{{\tilde F}^{\mu \nu }} \= {B^\mu }{u^\nu } - {B^\nu }{u^\mu } + \tilde F_{(1)}^{\mu \nu }, 
\end{eqnarray}
\end{subequations}
where we introduced a unit vector ${b^\mu } = {B^\mu }/\sqrt {{B^\nu }{B_\nu }}$ and the projection operator ${\Xi ^{\mu \nu }} = {\Delta ^{\mu \nu }} - {b^\mu }{b^\nu }$ 
such that $b_\mu \Xi ^{\mu \nu }=0= u_\mu \Xi ^{\mu \nu }$; 
Note that $ u_\mu B^\mu = 0$ by definition (\ref{eq:charges}). 
The explicitly written terms exhaust the zeroth-order terms 
that can be constructed with the available tensors 
in the absence of the vector and axial charges. 
The subscripts denote the first-order corrections 
that will be constrained by the entropy-current analysis below. 
{
Coefficients $p_{\para,\perp}$ are identified with thermodynamic quantities below. 
}

We are in position to compute the divergence of 
the entropy current $s^\mu = s u^\mu + s_\one^\mu $, 
where $s_\one^\mu $ is the first-order corrections 
to the entropy current. 
It can be expressed with 
the derivatives of the conserved quantities 
by the use of the first law of thermodynamics (\ref{con5}). 
Then, using the equations of motion (\ref{con4}) together 
with the constitutive equations (\ref{con6}), one finds that 
\begin{eqnarray}  \label{con9}
\pd_\mu s^\mu 
\= s \pd_\mu u^\mu + D s + \pd_\mu s_\one^\mu 
\\
\= \beta  ( Ts - \epsilon - p_\perp 
 + H_\mu B^\mu ) \pd_\mu u^\mu
\nnb
&&
- \beta \big[\, (p_\para - p_\perp)  b^\mu b^\nu 
+ B  b^\mu H^\nu  \, \big] \pd_\mu u_\nu 
- \Theta^{\mu\nu}_\one \pd_\mu\beta _\nu  
\nnb
&&
+  \tilde F^{\mu\nu}_\one \pd_\mu (\beta H_\nu )
 + \pd_\mu (s_\one^\mu + \beta u_\nu \Theta^{\mu\nu}_\one 
 - \beta H_\nu  \tilde F^{\mu\nu}_\one )
 \nn
.
\end{eqnarray}   
The leading-order terms in derivative describe 
the {\it ideal MHD}. 
For the entropy production to vanish 
in the ideal order, one should have 
\begin{subequations}
\label{con11}
\begin{eqnarray} 
&& p_\perp = p ,
\\
&&
({p_\parallel } - {p_ \bot }){b^\nu } + {H^\nu }B = 0
.
\end{eqnarray}
\end{subequations}
The zeroth-order result can be summarized as 
\begin{eqnarray} 
\Theta^{\mu\nu} _\zero 
\= \epsilon u^\mu u^\nu - p \Xi^{\mu\nu} 
+ ( p - B_\mu H^\mu) b^\mu b^\nu ,
\end{eqnarray} 
where $H^\mu = \mu_m^{-1} B^\mu $ 
with $\mu_m $ being the magnetic permeability. 
The above result indicates 
{that $p_\perp$ is identified with the thermodynamic pressure $p$ in Eq.~(\ref{con5}) and that} the pressure anisotropy is induced by the Maxwell stress.

The second law of thermodynamics requires that 
the first-order corrections be semipositive definite. 
This condition is satisfied if individual contributions 
of thermodynamic forces take semipositive values 
in Eq.~(\ref{con9}), i.e., 
\begin{subequations}
\label{eq:first-order}
\begin{eqnarray} 
&& 
- \Theta _{(1)}^{\mu \nu } {\partial _\mu }{\beta _\nu } \geq 0,
\\
&& 
  \tilde F_{(1)}^{\mu \nu } \pd_\mu (\beta {H_\nu }) \geq 0 
.
\end{eqnarray}
\end{subequations}
These inequalities can be ensured for 
general hydrodynamic configurations if 
the left-hand sides are organized in bilinear forms, 
constraining the possible forms of the constitutive equations as 
\begin{subequations}
\begin{eqnarray} 
\Theta_{(1)}^{\m\n} \= 
- T  \eta^{\m\n\r\s}  \pd_{(\r} \beta_{\s)}   
, 
\\
\tilde F_{(1)}^{\mu \nu } \= - T \rho ^{\mu \nu \rho \sigma }{\partial_ {[\rho }}(\beta {H_{\sigma ]}})
. 
\end{eqnarray} 
\end{subequations}
The fourth-rank tensors $ \eta^{\m\n\r\s}$ and $\rho ^{\mu \nu \rho \sigma } $ can be constructed with 
the available tensors, $ u^\mu$, $ b^\mu$, $ \Xi^{\mu\nu}$, 
and $\epsilon^{\mu\nu\a\b} $, as \cite{Grozdanov:2016tdf,Hongo:2020qpv, Hattori:2022hyo} 
\begin{subequations}
\begin{eqnarray}  
{\eta ^{\mu \nu \rho \sigma }}  
\= 
\begin{pmatrix}
b^\mu b^\nu & \Xi^{\mu\nu}
\end{pmatrix}
\begin{pmatrix}
\zeta_\para & \zeta_\times \\
\zeta'_\times & \zeta_\perp
\end{pmatrix}
\begin{pmatrix}
b^\rho b^\sigma \\ \Xi^{\rho\sigma}
\end{pmatrix}
\\
&&
+ 2{\eta _\parallel }({b^\mu }{\Xi ^{\nu (\rho }}{b^{\sigma )}} + {b^\nu }{\Xi ^{\mu (\rho }}{b^{\sigma )}})
 \nnb
 &&
 + {\eta _ \bot }({\Xi ^{\mu \rho }}{\Xi ^{\nu \sigma }} + {\Xi ^{\mu \sigma }}{\Xi ^{\nu \rho }} - {\Xi ^{\mu \nu }}{\Xi ^{\rho \sigma }}) 
 ,  \nnb
\rho ^{\mu \nu \rho \sigma} \= - 2{\rho_ \bot }({b^\mu }{\Xi ^{\nu [\rho }}{b^{\sigma ]}} - {b^\nu }{\Xi ^{\mu [\rho }}{b^{\sigma ]}})  +  2{\rho_\parallel }{\Xi ^{\mu [\rho }}{\Xi ^{\sigma ]\nu }}
.  \nnb
\end{eqnarray} 
\end{subequations}
{Plugging these expressions into Eq.~(\ref{eq:first-order}), one can confirm that all the terms on the left-hand sides are positive semi-definite bilinears. }
Note that we have chosen the Landau frame and 
the matching condition for the magnetic flux 
such that $e =  u_\mu u_\nu {\Theta_\zero ^{\mu \nu }} 
$ and $
B^\mu  = -  {\tilde F}_\zero^{\mu \nu } u_\nu  $ 
(see a review article \cite{Hattori:2022hyo} for more detailed  discussions). 
Therefore, the tensors $\eta ^{\mu \nu \rho \sigma } $ and $\rho ^{\mu \nu \rho \sigma } $ are transverse to the flow vector $ u^\mu $.  
It will be an interesting extension to discuss 
stability and causality in a general frame and matching conditions 
(see discussions in Sec.~\ref{sec:s-and-c} 
and recent works \cite{Bemfica:2017wps, Bemfica:2019knx, Bemfica:2020zjp, Kovtun:2019hdm, Hoult:2020eho, Armas:2022wvb}).

{
While the above tensor structure is somewhat involved, the physical meaning of the transport coefficients is clear (See \cite{Hongo:2020qpv, Hattori:2022hyo} for more details). 
Due to a preferred orientation specified by a magnetic field, the familiar shear viscosity $\eta$, bulk viscosity $\zeta$, and resistivity $\rho$ split into two components parallel and perpendicular to a magnetic field. 
The coefficient $\zeta_\times$ is a cross bulk viscosity that quantifies the pressure response along the parallel direction in response to compression in the perpendicular direction. 
The last one $\zeta'_\times$ is also a cross bulk viscosity with the directions swapped. 
}
Note also that we have $\zeta_\times = \zeta_\times' $ 
by virtue of Onsager's reciprocal relation and that 
there are no Hall terms in the charge-neutral systems. 

The second law of thermodynamics, i.e., the inequalities (\ref{eq:first-order}), basically requires 
all the transport coefficients introduced above be semipositive. 
An exception is the off-diagonal component $\zeta_\times $ that 
does not have to be semipositive definite 
since the second law can be ensured as long as 
the eigenvalues of the matrix are semipositive definite. 
In summary, one finds the inequalities 
\begin{eqnarray}
\label{inequality}
&&  {\zeta _ \bot } \ge 0, \quad 
{\zeta _\parallel } \ge 0, \quad 
{\zeta _\parallel }{\zeta _ \bot } \ge \zeta _ \times ^2,
\\
&&
 {\eta _\parallel } \ge 0,  \quad 
{\eta _ \bot } \ge 0,   \quad  
\rho_\perp \ge 0, \quad \rho_\parallel\ge 0
.   \nn
\end{eqnarray}

\section{Linear-mode analysis}\label{s3}

In this section, we solve the first-order hydrodynamic equations 
for the small perturbations near an equilibrium state, 
which is often called the linear-mode analysis.  
We apply perturbations on top of equilibrium values 
$u^\mu  = (1,0,0,0)$ and $B^\mu  = (0,0,0,B)$, 
where we took the direction of the magnetic field along the $ z$ axis at the equilibrium without loss of generality. 
Namely, the conserved charges are displaced from their equilibrium values as 
\begin{eqnarray}
&& e \to e+ \delta e (x), \ \ 
 {u^\mu } \to u^\mu  + \delta {u^\mu }(x) ,
\label{don01}
\nnb&&
{B^\mu } \to B^\mu  + \delta {B^\mu } (x)
. 
\end{eqnarray}
We will linearize the hydrodynamic equations with respect to these perturbations. 
The perturbation $\delta {B^\mu } $ can have 
a perpendicular component to $B^\mu $. 
We assume a linear relation 
$H^\mu= B^\mu / \mu_m $ with $ \mu_m$ 
being a constant in spacetime.  
For simplicity, we also assume that the contributions of 
the matter and magnetic components to the equilibrium 
energy density and pressure can be separated as 
\begin{eqnarray}
 p=P+\frac{B^2}{2\mu_m}, \quad 
e=\epsilon+\frac{B^2}{2\mu_m} .
\label{relations}
\end{eqnarray}

The conservation law of the energy-momentum tensor (\ref{con4}) 
can be projected as 
\begin{eqnarray} 
  {u_\nu }{\partial _\mu }{\Theta ^{\mu \nu }} = 0 , \quad 
  \Xi _\nu ^\rho {\partial _\mu }{\Theta ^{\mu \nu }} = 0 , \quad 
  {b_\nu }{\partial _\mu }{\Theta ^{\mu \nu }} = 0  .
\label{don04}
\end{eqnarray}
Plugging Eq.~(\ref{con6}) into the above 
and focusing on the linear order in the perturbations,  
one arrives at the linearized equations 
\begin{subequations}
\label{don13}
\begin{eqnarray} 
0 \=  \partial_0 \delta \epsilon+\frac{B}{\mu_m}\partial_0\delta B 
+h\partial_{\perp\mu} \delta u^\mu_\perp
+ (h-\frac{B^2}{\mu_m})  \partial_z \delta u_z,
\\
0\=h{\partial _0 }{\delta u_{x,y} } 
+ c_s^2{\partial _{x,y} }{\delta \epsilon }  
+\frac{B}{\mu_m} ( \partial_{x,y}\delta B_z -  {\partial_z }\delta {B_{x,y}}) 
\nnb
&&
-[(\zeta_\perp+\eta_\perp)\partial_{x,y}^2
+\eta_\perp\partial_{y,x}^2
+ \eta_\parallel \partial_z^2]\delta u_{x,y} 
\nnb
&&
- \zeta_\perp \partial_x\partial_y\delta u_{y,x}
-(\zeta_\times+\eta_\parallel )\partial_z\partial_{x,y} \delta u_z,
\\
0\=c_s^2{\partial _z }\delta \epsilon
+(h-\frac{B^2}{\mu_m}){\partial _0 }{\delta u_z } 
-(\zeta_\parallel\partial_z^2+ \eta_\parallel (\partial_x^2+\partial_y^2))\delta u_z
\nnb
&&
-(\zeta_\times+\eta_\parallel )\partial_z(\partial_x\delta u_x+\partial_y\delta u_y) 
,
\end{eqnarray}  
\end{subequations} 
where the subscripts $x,y,z $ denote the spatial components, 
but without minus signs from the metric, 
i.e., $(B_x,B_y,B_z) = (B^1,B^2,B^3) $, 
$(\pd_x,\pd_y,\pd_z) = (\pd^1,\pd^2,\pd^3) $, 
$  \delta u^\mu_\perp = (0,\delta u_x,\delta u_y,0)$. 
The second equation for the transverse components have 
the rotational symmetry around the magnetic-field direction. 
We also introduced the enthalpy $ h =   e +   p =
  \epsilon +   P +   B^2 /\mu_m $ with the equilibrium values 
and the (squared) sound velocity $c_s^2= \delta P/ \delta \epsilon$.

The equations for $\tilde F^{\mu\nu} $ can be 
projected and linearized in the same manner. 
The projected conservation law reads 
\begin{eqnarray} 
  {u_\nu }{\partial _\mu }{\tilde F^{\mu \nu }} = 0 , \quad 
  \Xi _\nu ^\rho {\partial _\mu }{\tilde F ^{\mu \nu }} = 0 , \quad 
  {b_\nu }{\partial _\mu }{\tilde F^{\mu \nu }} = 0  .
\label{eq:F-prj}
\end{eqnarray} 
The explicit forms of the linearized equations are obtained as 
\begin{subequations}\label{don26}
\begin{eqnarray} 
\label{eq:Gauss}
0 \= \partial_i\delta B^i , \\
0 \= B\partial_z \delta u_{x,y} -\partial_0 \delta B_{x,y}
\nnb
&& 
- \rho'_\perp T \big[ \, \partial_z\partial_{x,y} \delta (\beta B_z)-\partial_z^2 \delta(\beta B_{x,y}) \, \big] 
\nnb
&&
+ \rho'_\parallel T\big[ \,\partial_\bot^2 \delta(\beta B_{x,y})
-\partial_{x,y} \partial_{\perp\mu}\delta (\beta B_\perp^\mu) \, \big] ,
\\ 
0 \= -B\partial_{\perp\mu} \delta u^\mu_\perp-\partial_0 \delta B_z 
\nnb
&&
+ \rho'_\perp T\big[ \,\partial_\perp^2 \delta(\beta B_z)-\partial_z\partial_{\perp\mu} \delta (\beta B_\perp^\mu)\, \big]
,
\end{eqnarray}
\end{subequations}
where we used an identity $ 0 = (u_\mu + \delta u_\mu)
(B^\mu + \delta B^\mu ) = B \delta u_z - \delta B^0 + \order (\delta^2) $ and defined 
\begin{eqnarray}
    \rho'_\para = \frac{ \rho_\para}{\mu_m} , \quad 
     \rho'_\perp = \frac{ \rho_\perp}{\mu_m} .
\end{eqnarray} 
It is useful to notice that the set of equations (\ref{don26}) 
contains only two independent dynamical equations. 
The first equation (\ref{eq:Gauss}) does not contain a time derivative and is nothing but the Gauss law constraint. 
Another redundancy can be identified with 
an identity 
\begin{eqnarray}
\label{eq:sum-rule}
0 = \pd_\mu \pd_\nu \tilde F^{\mu\nu}
= (  \Xi_{\alpha\beta} 
-   u_\a   u_\b
+   b_\a    b_\b) \pd^\a \pd_\mu  \tilde F^{\mu\b}
. 
\end{eqnarray}
This identify is satisfied by any antisymmetric tensor 
regardless of the actual components of $\tilde F^{\mu\nu} $, 
and serves as a sum-rule constraint 
on the set of equations (\ref{don26}). 
Then, we are left with two independent dynamical equations 
and, correspondingly, the two spatial components of $\delta B^\mu $. 

\cout{[NOTE] $ 0 = \delta u_\mu B^\mu + u_\mu \delta B^\mu
= B \delta u_z - \delta B^0 $ and 
$0=(u_\nu + \delta u_\nu) \pd_\mu ( \tilde F_{\rm const}^{\mu\nu} 
+ \delta \tilde F^{\mu\nu} )
= u_\nu  \pd_\mu  \delta \tilde F^{\mu\nu} 
+  \delta u_\nu  \pd_\mu \tilde F_{\rm const}^{\mu\nu}  
+ \order(\delta^2)
=  u_\nu  \pd_\mu (\delta B^\mu u^\nu + B^\mu \delta u^\nu
- \delta B^\nu u^\mu + B^\nu \delta u^\mu ) 
= \pd_\mu (- \delta B^\mu -  u_\nu  \delta B^\nu u^\mu  ) 
= - \pd_\mu \delta B^\mu - \pd_t \delta B^0
= - \pd_i \delta B^i $}

The derivative of $ \delta \b$ in Eq.~(\ref{don26}) can be expressed with 
that of $\delta \epsilon $ with the help of 
a relation obtained from the thermodynamic relation (\ref{con5}), 
that is,  
\begin{eqnarray}
\label{eq:delta-b}
\delta\beta = -\frac{c_s^2 \beta}{h-  B^2/\mu_m}\delta\epsilon .
\end{eqnarray}

To summarize the above equations in the Fourier representation, 
we introduce a perturbation in a single mode 
\begin{eqnarray}
\label{eq:linear-solutions}
     \delta \bu (t,x,z) =  
     \delta \tilde \bu(\omega,k_\perp,k_\para) e^{- i \omega t + i k_\perp x + i k_\para z}
     ,
\end{eqnarray}
and the same for $\delta e $ and $\delta \bB $. 
Here, without loss of generality, 
we have set the transverse coordinate system in such a way that 
the dependence on the $y$ coordinate vanishes. 
Then, Eqs. (\ref{don13}) and (\ref{don26}) can be cast into two separate matrix equations 
\begin{subequations}
\label{eq:matrix-eqs}
\begin{eqnarray} 
\label{eq:Alfven-modes} 
\Big( \, A_\zero + i A_\one \, \Big) 
\begin{pmatrix}
  \delta B_y  \\ \delta u_y 
\end{pmatrix} 
\=0 ,
\\
\label{eq:sonic-modes} 
\Big( \, M_\zero + i M_\one \, \Big) 
\begin{pmatrix}
\delta \epsilon \\ \delta u_x \\  \delta u_z \\ \delta B_x \\ \delta B_z
\end{pmatrix}
\=0 .
\end{eqnarray}
\end{subequations}
The explicit forms of the first set of matrices are given as 
\begin{subequations} 
\label{eq:matrix-Alfven}
\begin{eqnarray}
A_\zero \= 
\begin{pmatrix}
 \omega   &  B k_\para 
\\
 h \frac{v_A^2}{B} k_\para &  h \omega 
\end{pmatrix} 
,  
\\
A_\one \=  
\begin{pmatrix}
  \rho_\perp' k_\para^2 + \rho_\para '  k_\perp^2  & 0
\\
 0  &  \eta_\para  k_\para^2 + \eta_\perp  k_\perp^2 
\end{pmatrix}
,
\end{eqnarray} 
\end{subequations}
with the so-called Alfv\'{e}n-wave velocity 
\begin{eqnarray}
\label{eq:Alfven-velocity}
v_A = \frac{B}{\sqrt{  \mu_m h }} .
\end{eqnarray}
The explicit forms of the second set of matrices are given as  
\begin{widetext}
\begin{subequations}
\label{eq:matrix-sonic}
\begin{eqnarray}  
\hspace{-1.2cm} 
M_\zero \=  
\begin{pmatrix} 
0 & - k_\perp  B & 0  & 0 & \omega   
\\
0 &  k_\para B & 0 &  \omega  & 0   
\\
\omega & -  h k_\perp &  
 h  ( v_A^2  - 1) k_\para & 0 & h \frac{v_A^2}{B}  \omega   
\\
- c_s^2 k_\perp &  h \omega & 0 &
h \frac{v_A^2}{ B} k_\para & - h \frac{v_A^2}{ B}   k_\perp   
\\
- c_s^2 k_\para & 0 &  -  h  ( v_A^2  - 1) \omega & 0 &  0 
\end{pmatrix} 
,
\\ \label{eq:M1}
\hspace{-1.2cm}
M_\one \= 
\begin{pmatrix} 
-   \frac{ B c_s^2 }{ h ( 1 -v_A^2) } 
\rho'_\perp  k_\perp^2
 & 0 & 0  & - \rho_\perp'  k_\para k_\perp  
&  \rho_\perp' k_\perp^2   
\\
\frac{ B c_s^2 }{ h ( 1 -v_A^2) } 
\rho'_\perp  k_\perp k_\para 
 & 0 & 0  &  \rho_\perp' k_\para^2 
& - \rho_\perp' k_\perp k_\para  
\\ 
0 & 0& 0 & 0& 0  
\\
0 & (\zeta_\perp + \eta_\perp  ) k_\perp^2 
+ \eta_\para k_\para^2 &
(\zeta_\times + \eta_\para  ) k_\perp k_\para &  0& 0 
\\
0 & (\zeta_\times + \eta_\para ) k_\perp k_\para &
 \zeta_\para k_\para^2 + \eta_\para k_\perp^2  &  0& 0  
\end{pmatrix}
.
\end{eqnarray}
\end{subequations}    
\end{widetext}
We will solve these equations in the next section. 
For later use, 
we introduce an angle $\theta $ measured from 
the direction of the magnetic field, and the momenta can be expressed as 
\begin{eqnarray}
k_\para = k \cos \theta , \quad 
k_\perp = k \sin \theta .
\end{eqnarray}
We also normalize all the viscous coefficients by the enthalpy, i.e., 
\begin{eqnarray}
\label{eq:viscosity-normalization}
\eta_{\para,\perp} '  
= \frac{1}{h} \eta_{\para,\perp}
, \quad 
\zeta_{\para,\perp,\times} '  
= \frac{1}{h} \zeta_{\para,\perp,\times}
.
\end{eqnarray}

\section{General solutions for the linearized equations}\label{s4}

In this section, we solve the matrix equations (\ref{eq:matrix-eqs}) to obtain the dispersion relations of the linear waves. 
The equations from the first-order hydrodynamics are accurate 
up to the order $ k^2$, so that our goal is to obtain 
the dispersion relation up to this order. 
We will obtain a complete set of analytic solutions with all the transport coefficients being free parameters. This is useful since the magnitudes of the transport coefficients are often not (precisely) known in individual systems.  
However, we also find that the small $k$ expansion poses an issue of convergence in anisotropic systems. 
We investigate the solutions near the angle $\theta\sim\pi/2$ in detail and provide an alternative series representation that works well in this regime.

We first discuss the analytic solutions for 
Eq.~(\ref{eq:Alfven-modes}), which have been discussed in the literature \cite{Grozdanov:2016tdf,Hernandez:2017mch, Armas:2022wvb}. 
We elaborate on this simpler equation to point out the convergence issue in the small $k$ expansion for anisotropic systems. 
We demonstrate the issue by comparing the limit of the angle $\theta\to\pi/2$ taken before and after the small $k$ expansion that does not agree with each other, and then identify the correct result, giving a different result than that in the literature \cite{Grozdanov:2016tdf, Armas:2022wvb}. 
We then provide a series representation that correctly captures this limit as well as the corrections in $k^2$ near this angle.

Then, we proceed to tackle the larger matrix in Eq.~(\ref{eq:sonic-modes}), of which the solutions have not been known in the literature. 
We will find the analytic solutions for the four modes fully including the dissipative effects. 
We introduce our simple algorithm for the solution search. 
We find that these modes also contain the convergence issue, 
and that the result at $\theta =\pi/2$ should be different from those in Refs.~\cite{Grozdanov:2016tdf, Armas:2022wvb}. 
An alternative series representation is provided accordingly.


\subsection{Alfv\'{e}n modes and issue of the small $k$ expansion in anisotropic systems}
\label{sec:Alfven}

The secular equation for Eq.~(\ref{eq:Alfven-modes}) 
is found to be 
\begin{eqnarray}
v_A^2  k_\para^2 
- (\omega + i \tilde \rho k^2 )( \omega + i \tilde \eta k^2)  = 0
,
\end{eqnarray}
where $\tilde \rho =  \rho'_\perp \cos^2 \theta
+\rho'_\para  \sin^2 \theta$ and $
\tilde \eta = \eta'_\para \cos^2 \theta + \eta'_\perp \sin^2 \theta  $. 
The solutions are readily obtained as 
\begin{eqnarray}
\label{eq:sol-Alfven}
\omega  \= \pm \sqrt{v_A^2  k_\para^2  
- \frac14  (\tilde \rho - \tilde \eta )^2k^4  }
- \frac{i}{2} (\tilde \rho + \tilde \eta) k^2
\nnb
\= \pm v_A k_\para 
- \frac{i}{2} (\tilde \rho + \tilde \eta) k^2 + \order (k^3)
.
\end{eqnarray}
We performed the small $k$ expansion in the second line. 
These solutions are gapless in the limit $k \to0 $ 
and are known as the Alfv\'{e}n waves propagating along the equilibrium magnetic field. 
Since $\eta_{\para,\perp} \geq0 , \, \rho_{\para,\perp} \geq0 $, 
these modes are damped out in time by an exponential factor 
$e^{- \frac12(\tilde \rho + \tilde \eta)t} $. 
Without a parity-breaking effect, we have a pair of waves propagating in opposite directions with the same damping rate.

Below, we elaborate on an issue of 
the small $k$ expansion involved in the Alfv\'{e}n modes. 
It is important to clarify this issue here because one will find the same issue in the other matrix equation (\ref{eq:sonic-modes}) of which the analytic solutions have not been known. 
Anisotropic systems may potentially share the same issue. 
We investigate the limit of angle $\theta \to \pi/2$, i.e., the vanishing $k_\para $ limit in Eq.~(\ref{eq:sol-Alfven}). 
Taking the limit {\it without} performing the small $k$ expansion, one finds that 
\begin{eqnarray}
\label{eq:sol-Alfven-limit}
\omega(\theta=\frac{\pi}{2})  \= \pm \frac{i}{2} 
|\tilde \rho - \tilde \eta | k_\perp^2 
- \frac{i}{2} (\tilde \rho + \tilde \eta) k_\perp^2
\nnb
\=  - i \eta'_\perp  k_\perp^2  , \ \  
- i \rho'_\para k_\perp^2 
,
\end{eqnarray}
irrespective of the sign of $\tilde \rho - \tilde \eta$. 
In this limit, these modes split into two distinct purely diffusive modes. 
These two modes are still invariant under the parity transformation, i.e., $k_\perp \to - k_\perp$, because the linear term vanishes in this limit. 
One can trace back the splitting of the dispersion relations (\ref{eq:sol-Alfven-limit}) to the original matrices (\ref{eq:matrix-Alfven}). 
Taking the limit $\theta \to \pi/2$, 
one finds that the matrix equation (\ref{eq:Alfven-modes}) reduces to a diagonal form  
\begin{eqnarray}
\label{eq:Alfven-diagonal}
\begin{pmatrix}
 \omega  +i \rho_\para '  k_\perp^2  &  0
\\
0 &   \omega  + i  \eta'_\perp  k_\perp^2 
\end{pmatrix} 
\begin{pmatrix}
  \delta B_y  \\ \delta u_y 
\end{pmatrix} 
=0 ,
\end{eqnarray}
where the two perturbations $  \delta B_y $ and $ \delta u_y $ are decoupled from each other. 
One of the dispersion relations (\ref{eq:sol-Alfven-limit}) is for the flow perturbation damped by the shear viscosity, while the other is for the magnetic flux diffusion by the resistivity.

Now, it should be noticed that the dispersion relations (\ref{eq:sol-Alfven-limit}) are not reproduced by the limit taken {\it after} performing the small $k$ expansion in Eq.~(\ref{eq:sol-Alfven}); The expanded result instead yields two degenerate purely diffusive modes \cite{Grozdanov:2016tdf, Armas:2022wvb}. 
This disagreement occurs due to an invalid expansion of the terms containing $k_\para = k \cos \theta$ that is not a small quantity but is exactly zero when $\theta \to \pi/2$. 
Performing the small $k$ expansion first and then taking the limit $\theta \to \pi/2$, one finds that 
\begin{eqnarray}
    \omega (\theta \to \frac{\pi}{2}) \= 
- \frac{i}{2} (  \rho'_\para + \eta'_\perp) k_\perp^2 
\nnb&&
\pm \lim_{\theta \to \pi/2} \sum_{n=1}^\infty 
\frac{c_n (\rho'_\para - \eta'_\perp )^{2n}}{ (v_A \cos \theta)^{2n-1} }  k_\perp^{2n+1} 
,
\end{eqnarray}
where $c_n$ is the numerical coefficients. 
In the above expansion, one encounters divergence of the higher-order terms as $\theta \to \pi/2$, which spoils the small $k$ expansion near $\theta = \pi/2$. 
Clearly, the small $k$ expansion and the limit of $\theta \to \pi/2$ do not commute with each other.  

There is a transient angle $\theta$ (for a given $k$) where the propagating modes turn into the purely diffusive modes. 
We investigate this transition below. 
The disagreement about the limits originates from the ill-organized small $k$ expansion when there is another small quantity $\cos \theta$. 
In this case, we should specify which of 
$k$ or $\cos\theta$ is smaller before carrying out an expansion. {In principle, one needs to make a dimensionless expansion parameter $\hat{k}\equiv k/k_c$ with an ultraviolet (UV) cutoff $k_c$ in order to compare the two small expansion parameters. 
While $k_c$ is implicit in the transport coefficients, it is useful to explicitly introduce $k_c$ so that one can maintain general values of the transport coefficients.} 
The cutoff in general depends on details of the microscopic dynamics, or more precisely, how we integrate out the UV degrees of freedom. 
Then, one can organize two pairs of series representations: 
\begin{subequations}
\label{eq:Alfven-various-expansion}
    \begin{eqnarray}
\omega \= \pm v_A k_\para 
- \frac{i}{2} (\tilde \rho + \tilde \eta) k^2 + \order \big(\frac{k}{|\cos\theta|} k^2 \big) 
,
\nnb
&&
{\rm for} \quad 
\hat{k} < |\cos\theta| ,
\\
\omega  \= - i k^2 \rho'_\parallel
- i \Big( ( \rho'_\perp - \rho'_\para) k^2
+ \frac{v_A^2}{\eta'_\perp - \rho'_\para} \Big)
\cos^2\theta
\nnb
&&
+\order(\cos^4\theta) , 
\nnb
&&
- i k^2 \eta'_\perp
- i \Big( (\eta'_\para - \eta'_\perp) k^2
-\frac{ v_A^2}{\eta'_\perp- \rho'_\para}\Big)\cos^2\theta
\nnb
&&
+\order(\cos^4\theta)
,\quad 
\nnb
&&
{\rm for} \quad 
|\cos\theta| < \hat{k} 
.
    \end{eqnarray}
\end{subequations}
While the first expression is the same as the expansion in the second line of Eq.~(\ref{eq:sol-Alfven}), it should be emphasized that the correction terms are small only when $\hat{k}/|\cos\theta|<1$. 
When $ |\cos\theta| < \hat{k} $, we find two expansions with distinct pure imaginary coefficients shown in the second and third lines. They are smoothly connected to the two purely diffusive modes (\ref{eq:sol-Alfven-limit}) at $\theta = \pi/2$. 
The inverse factor of $(\eta'_\perp- \rho'_\para)$ does not cause a divergence in general unless fine-tuned.

\cout{Here we introduce an ultraviolet (UV) cutoff $k_c$ to make a dimensionless expansion parameter $\hat{k}\equiv k/k_c$. The cutoff depends on the microscopic dynamic details, or more precisely, how we integrate out the UV degrees of freedom. In order to compare the two small expansion parameters, the presence of $k_c$ seems to be necessary. However, $k_c$ is not expected to appear in the hydrodynamic dispersion relations  because it can be totally  absorbed into transport coefficients which are left unspecified within the present discussion. 
On the other hand, what concerns us is how small $\hat{k}$ is compared to $\cos\theta$, therefore, it really counts how small $k$ is with $k_c$ fixed after choosing a specific coarse-grained procedure. In other words, there is nothing to bother about the introduction of $k_c$.
}

\begin{figure}[t]
    \centering
    \includegraphics[width=\hsize]{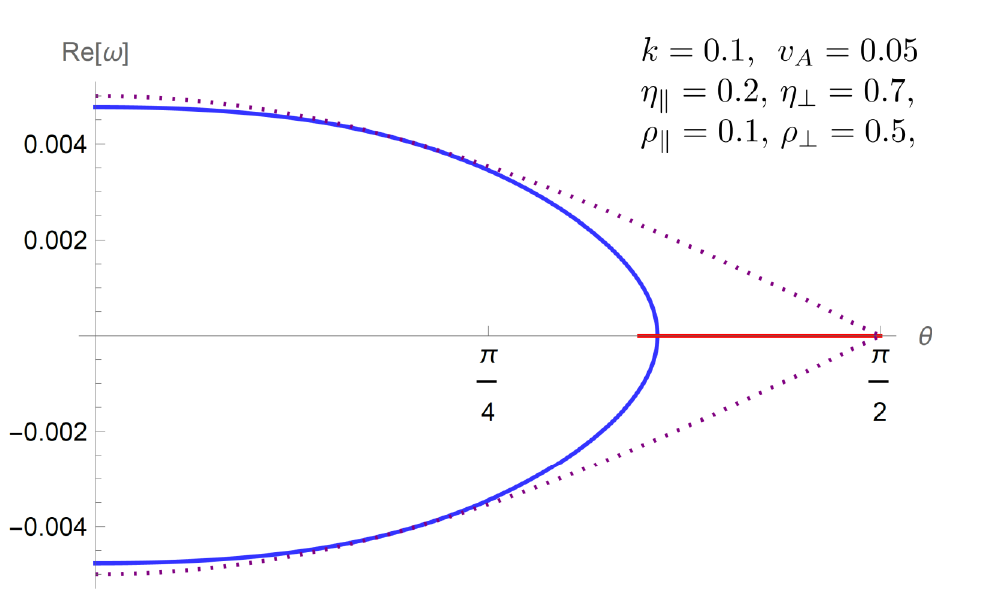}
    \\
        \includegraphics[width=\hsize]{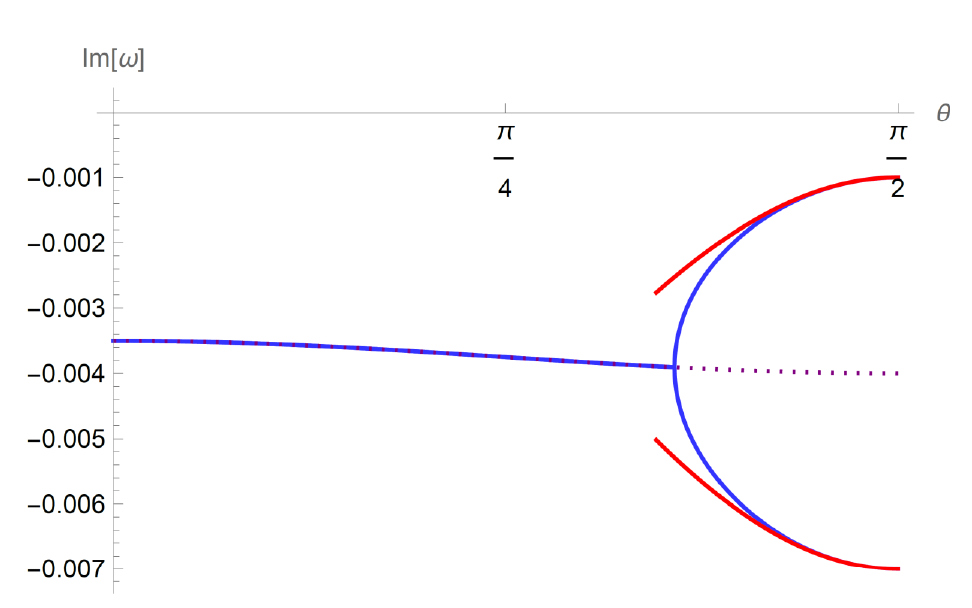}
    \caption{The dispersion relation of the Alfv\'{e}n modes. The original solution (blue-solid curves) in Eq.~(\ref{eq:sol-Alfven}) is compared with the small-$k$ ({dotted lines}) and small-cosine expansions ({red lines}) in Eq.~(\ref{eq:Alfven-various-expansion}). 
    The parameters are fixed as indicated in the legend for illustration. 
    } 
    \label{fig:Alfven}
\end{figure}

In Fig.~\ref{fig:Alfven}, we plot the above series representations together with the original solution shown in Eq.~(\ref{eq:sol-Alfven}). 
We confirm the agreement between the small-cosine expansion (red-solid curves) and the original solution (blue-solid curves) for the angle near $\theta = \pi/2$. As noted above, the small $k$ expansion (dotted curves) breaks down as the angle approaches $\theta = \pi/2$. 
There is a critical angle where the real part, and thus the velocity, vanishes. 
The critical angle is simply determined by the condition that the square root vanishes in Eq.~(\ref{eq:sol-Alfven}). 
Above the critical angle, the Alfv\'{e}n modes turn into purely diffusive modes, and the degenerate imaginary parts split into two distinct values. 


\cout{
\begin{subequations}
\label{eq:Alfven-various-expansion}
    \begin{eqnarray}
\omega \= \pm v_A k_\para 
- \frac{i}{2} (\tilde \rho + \tilde \eta) k^2 + \pp{ \order \Big(\frac{\hat{k}}{|\cos\theta|}\hat{k}^2 \Big) 
, \quad \hat{k} < |\cos\theta| < 1,
}
\\
\omega 
\= \pm \pp{k_\para} \sqrt{v_A^2   
- \frac14 \pp{ k_{\rm cut}^2 } (\tilde \rho - \tilde \eta )^2 } 
- \frac{i}{2} (\tilde \rho + \tilde \eta) k^2,\quad \hat{k}=|\cos\theta| <1,
\\   
\omega  \= \cgd{-{i k^2 \rho_\parallel}+i({ k^2(\rho_\para-\rho_\perp)}-\frac{h v_A^2}{\eta_\perp-h \rho_\para})\cos^2\theta+\order(cos^4\theta)}
\\
\= \cgd{-\frac{i k^2 \eta_\perp}{h}+i(\frac{ k^2(\eta_\perp-\eta_\para)}{h}+\frac{h v_A^2}{\eta_\perp-h \rho_\para})\cos^2\theta+\order(cos^4\theta)}
,\quad |\cos\theta| < \hat{k} < 1
.
    \end{eqnarray}
\end{subequations}  
The magnitude of $\cos \theta$ becomes smaller from the first to the third expressions relative to $\hat k$, of which the leading-order terms agree with those in Eq.~(\ref{eq:sol-Alfven}) and (\ref{eq:sol-Alfven-limit}), respectively. 
It is interesting to see the marginal case in the middle where we used $k^2=k_\para k_{\rm cut}$.  
that connects the propagating case with the purely diffusive case. Depending on the sign within the square root, this mode will merge into the propagating region or the diffusive region.
When $\theta$ is zero and $\hat{k}=|\cos\theta| \to 1$, this marginal situation corresponds to a ``critical" case because hydrodynamic expansion breaks down at that point.

\com{[Note]
$k_\para \sim k \cdot \frac{k}{k_c} = \frac{k^2}{k_c}$
\begin{eqnarray}
    \omega  \= \pm \sqrt{v_A^2 k_\para^2 
- \frac14  (\tilde \rho - \tilde \eta )^2 (k_\para k_c)^2 }
- \frac{i}{2} (\tilde \rho + \tilde \eta) k^2
\nnb 
\= \pm k_\para \sqrt{v_A^2  
- \frac14 k_c^2 (\tilde \rho - \tilde \eta )^2  }
- \frac{i}{2} (\tilde \rho + \tilde \eta) k^2
\\
    \omega  \= \pm \sqrt{v_A^2\frac{k^4}{k_c^2}
- \frac14  (\tilde \rho - \tilde \eta )^2k^4 }
- \frac{i}{2} (\tilde \rho + \tilde \eta) k^2\nonumber\\
\= \pm k^2 \sqrt{v_A^2  k_c^{-2}  
- \frac14  (\tilde \rho - \tilde \eta )^2 }
- \frac{i}{2} (\tilde \rho + \tilde \eta) k^2\nonumber\\
\= \pm \sqrt{v_A^2  k_c^{-2}  
- \frac14  (\tilde \rho - \tilde \eta )^2 }k_c |\cos\theta|k
- \frac{i}{2} (\tilde \rho + \tilde \eta) k^2,\quad \hat{k}=|\cos\theta| < 1.
\end{eqnarray}
}

}

\subsection{magnetosonic modes from analytic algorithm}

\label{sec:magneto-sonic}

In the previous subsection, we investigated the Alfv\'{e}n modes encoded in the matrix equation (\ref{eq:Alfven-modes}). 
The analytic solutions for the other matrix equation (\ref{eq:sonic-modes}) have not been known to the best of 
our knowledge, except for the solutions at the particular momentum directions, 
i.e., $ k_\para = 0$ or $k_\perp =0 $ \cite{Grozdanov:2016tdf, Hernandez:2017mch, Armas:2022wvb}. 
It is challenging to obtain analytic solutions including the higher-order corrections in $k $. 
If one invokes brute-force efforts, 
one has to find general solutions for a quartic equation 
in $\omega $, which is possible 
but is not an efficient path to reach compact forms of solutions. 
Moreover, the solutions at $ k_\para = 0$ suffer from the issue of the small $k$ expansion discussed in the previous subsection, giving rise to different solutions in this limit than those obtained in Refs.~\cite{Grozdanov:2016tdf, Armas:2022wvb}.
As in the Alfv\'{e}n modes, we investigate the behavior near this limit carefully.

First, we provide a simple method for the solution search 
that works on an order-by-order basis in $ k$. 
This method serves as a general algorithm 
that can be applied to general sets of hydrodynamic equations and any other equations based on derivative expansions (see Ref.~\cite{Fang:2024sym} for further applications), while we here focus on the quartic secular equation in $\omega $ at $k^2 $ order.  
We note again that the secular equations for hydrodynamic equations, and thus their solutions, are only accurate up to a given order in $k$. 
Thus, the order-by-order algorithm is a suitable method for the solution search.

We begin with the leading-order solutions 
by putting $M_\one =0 $. 
The secular equation from $M_\zero$ is found to be 
\begin{eqnarray}
\omega \Big[ \omega^4 -  \V^2 k^2 \omega^2
+ v_A^2 c_s^2 k^2 k_\para^2 \Big] = 0
,
\end{eqnarray}
with $ \V^2 := (c_s^2 +v_A^2) - c_s^2 v_A^2 \sin^2 \theta \geq 0$. 
A trivial solution, $ \omega =0$, originates from the redundancy 
mentioned below Eq.~(\ref{eq:sum-rule}). 
We exclude this trivial solution in the following discussions. 
The leading-order dispersion relations are found to be 
\begin{eqnarray}
\omega = \pm v_1 k, \quad  \pm v_2 k
,
\end{eqnarray}
where the two distinct velocities are given as 
\begin{eqnarray}
\label{eq:sonic-velocities}
v_{1,2} = \frac{1}{\sqrt{2}} 
\sqrt{  \V^2 \pm \sqrt{ \V^4 - 4 v_A^2 c_s^2 \cos^2 \theta } }
 \ ,
\end{eqnarray} 
where the upper and lower signs are for $v_1 $ and $ v_2$, respectively. 
These modes are two pairs of counterpropagating waves 
called the fast and slow magnetosonic waves.
When $\theta=0$, one of the pairs reduces to the Alfv\'{e}n waves (\ref{eq:sol-Alfven}) as required by the rotational symmetry and the other pair reduces to the sound modes without modification of the sound velocity because of the absence of a magnetic pressure according to the Gauss law $\delta B_z = - k_\perp \delta B_x/k_\para = 0$. 
In this limit, one finds that 
$\{v_1, v_2\} = \{c_s, v_A\}$ when $c_s\geq v_A$ 
and $\{v_1, v_2\} = \{v_A, c_s\}$ when $c_s < v_A$.

It is important to note that 
a pair of counterpropagating waves acquires 
the same dissipative corrections at $k^2 $ order 
in the absence of parity-breaking effects. 
Therefore, the general solutions should be found in the forms 
\begin{eqnarray}
\label{eq:sonic-solutions}
\omega = \pm v_1 k - i w_1 k^2  , \quad 
\omega = \pm v_2 k - i w_2 k^2 ,
\end{eqnarray}
where $w_{1,2} $ are independent of $ k$ 
and are determined below. 
Accordingly, one can make an ansatz for 
the factorized form of the quartic secular equation as 
\begin{eqnarray}
\label{eq:ansatz}
f_{\rm ans} (\omega) \= 
\{ \omega - (v_1 k - i w_1 k^2) \} 
\{ \omega - (-v_1 k - i w_1 k^2) \} 
\nnb
&& \times
\{ \omega - (v_2 k - i w_2 k^2) \} 
\{ \omega - (-v_2 k - i w_2 k^2) \} 
\nnb
&&
+ \order(\omega^3 k^3)
+ \order(\omega^2 k^{4})
+ \order(\omega^1 k^{5})
+ \order(\omega^0 k^{6})
. \nnb
\end{eqnarray} 
The uncertainties at $k^3 $ order in each solution 
result in the uncertainties indicated in the last line. 
These uncertainties are not of our interest here, 
since they are not improved unless 
the constitutive equations are improved beyond 
the first-order derivative expansions.  
The following computation is greatly simplified by identifying these irrelevant higher-order terms and getting rid of them at this stage. 
Expanding $f_{\rm ans} (\omega) $, 
we should only retain the relevant terms as 
\begin{eqnarray}
\label{eq:ansatz-exp}
\hspace{-0.5cm}
f_{\rm ans} (\omega) = \omega^4 
+ g_2(k) \omega^3 + g_3(k) \omega^2 + g_4(k) \omega + g_5 (k)
 , 
\end{eqnarray}
where $g_n $ denotes the $n $-th polynomial of $k $ stemming from the expansion of Eq.~(\ref{eq:ansatz}). 
As mentioned above, we should not, or do not have to, retain the terms higher than $n $ that could only be relevant beyond the first-order hydrodynamics.

The ansatz (\ref{eq:ansatz-exp}) is matched to 
the secular equation from Eq.~(\ref{eq:sonic-modes}). 
Consistent to the above order counting, 
we only need to retain the terms at the same orders in $ k$ 
in the secular equation. 
Then, the matching for the coefficients 
in the $\omega^3 $ and $\omega^1 $ terms lead to coupled linear equations 
\begin{subequations}
\label{eq:sonic-small-k}
\begin{eqnarray} 
&&
v_2^2 w_1 + v_1^2 w_2 = \frac12 W_1  
, 
\\
&&
w_1 + w_2  = \frac12 W_2
  ,
\end{eqnarray}
\end{subequations}
The explicit forms of $W_{1,2} $ are given as 
\begin{subequations}
\label{eq:W-sol}
\begin{eqnarray} 
W_1  \= \eta'_\para \Big( c_s^2  \cos^2 (2\theta)
+  \frac{v_A^2  \sin^2 \theta}{1-v_A^2}
 \Big) 
 \\
 &&
+ \rho'_\perp c_s^2 \frac{ 1 - v_A^2 \cos^2 \theta }{1-v_A^2}  
+ 
\zeta'_\para   \frac{v_A^2}{1-v_A^2}  \cos^2 \theta 
\nnb
&&
+ \big\{ \zeta'_\para - 2 \zeta'_\times 
+ (\zeta'_\perp + \eta'_\perp) \big \} c_s^2  \sin^2 \theta \cos^2 \theta 
,  
\\
W_2 \=  \eta'_\para  \Big( 1 + \frac{v_A^2}{1-v_A^2} \sin^2 \theta\Big)  
+ \rho'_\perp \Big( 1 + \frac{v_A^2  c_s^2 }{1-v_A^2} \sin^2\theta  \Big)
\nnb
&&
+  \frac{\zeta'_\para}{1-v_A^2} \cos^2 \theta 
+ \big( \zeta'_\perp  + \eta'_\perp ) \sin^2\theta  
,
\end{eqnarray}
\end{subequations}
where all the viscous coefficients are normalized by the enthalpy as in Eq.~(\ref{eq:viscosity-normalization}). 
Note that the coefficients in the $\omega^2 $ and $\omega^0 $ terms are automatically matched when one inserts $ v_{1,2}$ in the leading-order solutions (\ref{eq:sonic-velocities}) because the unknowns $w_{1,2} $ are not involved. 
It is now a quite simple task to solve the above linear equations 
to find the solutions 
\begin{eqnarray} 
\label{eq:ws}
w_1 = - \frac{ W_1 - v_1^2 W_2 }{2(v_1^2 - v_2^2)}  
, \quad 
w_2 = + \frac{ W_1 - v_2^2 W_2 }{2(v_1^2 - v_2^2)}  
.
\end{eqnarray} 
As expected in the ansatz (\ref{eq:ansatz}), $w_1 $ and $w_2 $ 
are interchanged when we interchange $v_{1} $ and $v_2 $. 
The simple algorithm leading to these analytic solutions can be applied to 
general equations based on derivative expansions 
even with higher-order terms in $\omega $ and/or $ k$. 

\cout{
In the above, we have obtained the analytic solutions for Eq.~(\ref{eq:sonic-modes}) in the forms of Eq.~(\ref{eq:sonic-solutions}) with the coefficients shown in Eqs.~(\ref{eq:sonic-velocities}) and (\ref{eq:ws}), 
These solutions have been missing in the literature for a long time.  
We emphasize that the simple algorithm leading to these analytic solutions can be applied to 
general equations based on derivative expansions 
even with higher-order terms in $\omega $ and/or $ k$. 
}

Now, making the use of the lesson from the Alfv\'{e}n waves discussed in Eq.~\ref{sec:Alfven}, we point out that the magnetosonic modes also suffer from the breakdown of the small $k$ expansion near $\theta=\pi/2$. This is again caused by another small quantity, $\cos \theta$, that induces divergence in the higher-order terms in the small $k$ expansion. 
When the cosine factor becomes small near $\theta=\pi/2$, one should use the small cosine expansion to get a correct result. 
To see this issue, we first take the limit  $\theta \to \pi/2 $ in the velocities (\ref{eq:sonic-velocities}). 
In this limit, one finds that 
\begin{eqnarray}
\label{eq:sonic-limit}
v_{1}(\theta =\frac{\pi}{2} ) = \sqrt{ c_s^2 +v_A^2 - c_s^2 v_A^2 } , \quad 
v_{2}(\theta =\frac{\pi}{2} )  = 0  .
\end{eqnarray} 
The slow magnetosonic waves do not propagate 
in the perpendicular direction and become purely diffusive modes. 
This implies the potential occurrence of the issue because, if there were a linear term, the counter-propagating modes should have a degenerate damping rate because of the parity invariance. 
The damping rates $w_{1,2} $ in the same limit read 
\begin{subequations}
\label{eq:sonic-damping-perp}
\begin{eqnarray} 
w_1 (\theta =\frac{\pi}{2} )  
\= \frac12 \Big( 
\eta'_\perp + \zeta'_\perp 
+ \rho'_\perp (1-c_s^2)^2 \frac{  v_A^2}{v_1^2}
\Big) ,
\label{eq:w1-limit}
\\
w_2 (\theta =\frac{\pi}{2} ) 
&\overset{?}{=}&\frac{ \eta'_\para}{2 (1-v_A^2)}
+ \frac{ \rho'_\perp c_s^2 }{2 v_1^2 (1-v_A^2)}
\label{eq:w2-limit}
.
\end{eqnarray}
\end{subequations} 
Just below, we confirm that the above degenerate $w_2 (\theta =\frac{\pi}{2} ) $ is {\it not} a correct result. 
To get the correct result, we take the limit $\theta \to \pi/2$ in Eq.~(\ref{eq:matrix-sonic}). Then, the matrix equation reads    
\begin{widetext}
\begin{eqnarray}
\label{equation_pi/2}
\hspace{-0.8cm}
\begin{pmatrix} 
-  i \frac{ B c_s^2  }{ h( 1 -v_A^2) } \rho'_\perp k_\perp^2 &
- k_\perp B   & \omega + i\rho'_\perp  k_\perp^2   & 0 
\\
\omega & -  h k_\perp  & h \frac{ v_A^2}{B} \omega  & 0 
\\
- c_s^2 k_\perp &  h \omega 
+i  (\zeta_\perp + \eta_\perp  ) k_\perp^2 
 & - h \frac{ v_A^2}{B}  k_\perp & 0 
\\
0 & 0 & 0  & h  ( 1-v_A^2) \omega 
+ i\eta_\para   k_\perp^2  
\end{pmatrix} 
\begin{pmatrix}
\delta \epsilon \\ \delta u_x \\ \delta B_z  \\ \delta u_z
\end{pmatrix}
= 0
,
\end{eqnarray}    
\end{widetext}
where we used $\delta B_x = - k_\para \delta B_z/k_\perp =0$ 
from the Gauss law (\ref{eq:Gauss}). 
Similar to the case of the Alfv\'{e}n modes (\ref{eq:Alfven-diagonal}), one readily finds decoupling of a flow perturbation $\delta u_z$ of which the dispersion relation is solely governed by the shear viscosity $\eta_\para$. 
Diagonalizing the remaining three modes and retaining the terms in the $k^2$ order, the dispersion relations at $\theta=\pi/2$ are found to be 
\begin{subequations}
\label{eq:sonic-disp-limit}
 \begin{eqnarray}
\label{eq:sonic-disp-limit-12}
\omega_{1,2} (\theta=\frac{\pi}{2}) \=  \pm v_1 k_\perp
- \frac{i}{2} \Big( 
\eta'_\perp + \zeta'_\perp 
+ \rho'_\perp (1-c_s^2)^2 \frac{  v_A^2}{v_1^2}
\Big)  k_\perp^2 
\nnb
&&
+ \order(k_\perp^3) 
,
\\
\label{eq:sonic-disp-limit-3}
\omega_3 (\theta=\frac{\pi}{2}) \= - \frac{i \rho'_\perp c_s^2 }{ v_1^2 (1-v_A^2)} k_\perp^2  + \order(k_\perp^4)
, 
\\
\label{eq:sonic-disp-limit-4}
\omega_4 (\theta=\frac{\pi}{2})\= - \frac{ i \eta'_\para}{1-v_A^2}  k_\perp^2 + \order(k_\perp^4) 
.
\end{eqnarray}   
\end{subequations}
Here, $v_1 = v_1(\theta=\pi/2)$ in Eq.~(\ref{eq:sonic-limit}) is understood. 
The fast magnetosonic modes (\ref{eq:sonic-disp-limit-12}) remain propagating modes, and still have the degenerate damping rate that agrees with $w_1 (\theta =\pi/2 ) $ in Eq.~(\ref{eq:w1-limit}). 
In contrast, the slow magnetosonic modes reduce to the two purely diffusive modes, and the damping rates split into two distinct ones that only depend on either $\eta'_\para$ or $\rho'_\perp$. 
They are different from the degenerate damping rate $w_2 (\theta = \pi/2 ) $ in Eq.~(\ref{eq:w2-limit}) that was  shown in Refs.~\cite{Grozdanov:2016tdf, Armas:2022wvb}.\footnote{ 
The same results as in Eqs.~(\ref{eq:sonic-disp-limit-3}) and (\ref{eq:sonic-disp-limit-4}) are shown in Ref.~\cite{Hernandez:2017mch}, where the limits are taken for $\theta\to\pi/2$ first and then $k\to0$. }


Now, we investigate the behaviors near $\theta = \pi/2$. 
When $\cos\theta<\hat{k}$, one should organize a series representation with respect to $\cos\theta$, as we have discussed in Eq.~(\ref{eq:Alfven-various-expansion}). 
To find the solutions in the series representations, one can apply the same algorithm introduced above. 
Then, we find the solutions in the form 
\begin{eqnarray} 
\label{eq:sonic-small-cos}
\omega_i(\cos\theta)\= \tilde \omega_i-\frac{\tilde\omega_i^3 X_3+\tilde\omega_i^2 X_2+\tilde\omega_i X_1+X_0}{\prod_{i\neq j}(\tilde\omega_i-\tilde\omega_j)}\cos^2\theta
\nnb
&& +\order(\cos^4\theta), 
\end{eqnarray} 
where $\tilde \omega_i$ on the right-hand side are the solutions for Eq.~($\ref{equation_pi/2}$) at $\cos^2\theta=0$, i.e., $\tilde \omega = \omega(\theta = \pi/2)$. 
It is interesting that the fast and slow sonic modes, which were previously labeled as $\omega_{1,2}$ and $\omega_{3,4}$, are mixed among themselves in the correction terms of order $\cos^2 \theta$. 
The explicit forms of $X_i$ are given as   
\begin{widetext}
\begin{subequations}
\label{eq:}
\begin{eqnarray} 
X_0 \=  c_s^2 v_A^2  k^4 
+ k^6  c_s^2 \rho'_\perp  \Big[ 
 ( \zeta'_\perp +\eta'_\perp)
 - \frac{2 \zeta'_\times  }{1-v_A^2} 
 + \frac{\zeta'_\para - 2(2-v_A^2) \eta'_\para}{(1-v_A^2)^2} 
\Big]
,
\\
X_1 \=  i k ^4 \Big[
\frac{ v_A^2 (\eta'_\para - \zeta'_\para + c_s^2 \rho'_\perp)}{ 1 -v_A^2} 
- c_s^2 (\zeta'_\para + \zeta'_\perp -2 \zeta'_\times - 4 \eta'_\para + \eta'_\perp )
\Big]
\\
&& 
+ \frac{ ik^6 }{ (1-v_A^2)^2} 
\rho'_\perp \Big[ c_s^2 v_A^2 \big(
\zeta_\times^{\prime\,2} + 2 \zeta'_\times \eta'_\para 
- \zeta'_\para ( \zeta'_\perp + \eta'_\perp)
+ 3 \eta'_\para (\zeta'_\perp + \eta'_\perp) 
\big) 
\nnb
&&
+ (1-v_A^2) \big( \zeta_\times^{\prime\,2} - \zeta'_\para(\zeta'_\perp + \eta'_\perp )
+ 2 \eta'_\para (\zeta'_\perp + \zeta'_\times + \eta'_\perp)
\big)
\Big]  
\nn
,
\\
X_2\= - c_s^2 v_A^2 k^2 - \frac{k^4}{(1-v_A^2)^2}
\Big[  c_s^2 v_A^2 \rho'_\perp \big( 
\zeta'_\para - 2 (1-v_A^2) \zeta'_\perp - (1+v_A^2) \eta'_\para 
-2 (1-v_A^2) \eta'_\perp \big)
\nnb
&&
- (1-v_A^2) \Big\{ 
\rho'_\perp \big(\, v_A^2 \eta'_\para 
+  (1-v_A^2)( \zeta'_\perp + \eta'_\perp) \, \big)
+ \zeta_\times^{\prime \,2} + 2 \eta'_\para (\zeta'_\perp+ \zeta'_\times + \eta'_\perp) 
- \zeta'_\para (\zeta'_\perp + \eta'_\perp + \rho'_\perp)
\Big\}
\Big]
,
\\
X_3\=  \frac{i k^2 }{1-v_A^2} \Big[ 
 \zeta'_\para - (1-v_A^2) (\zeta'_\perp + \eta'_\perp)
- v_A^2 ( \eta'_\para + c_s^2  \rho'_\perp )
\Big]
,
\end{eqnarray}
\end{subequations}  
\end{widetext}
These results should replace the naive small $k$ expansion when $\cos\theta<\hat{k}$.

\cout{
\begin{figure}
    \centering
    \includegraphics[width=\hsize]{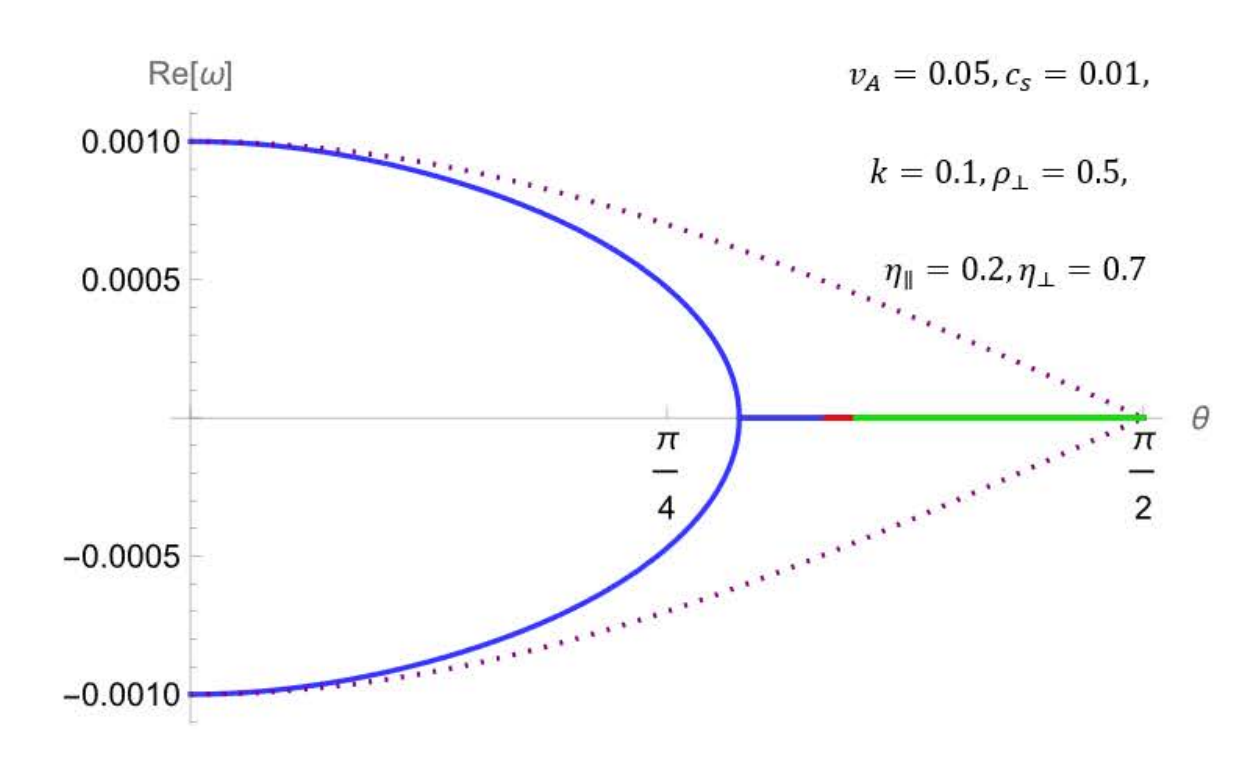}
    \\
        \includegraphics[width=\hsize]{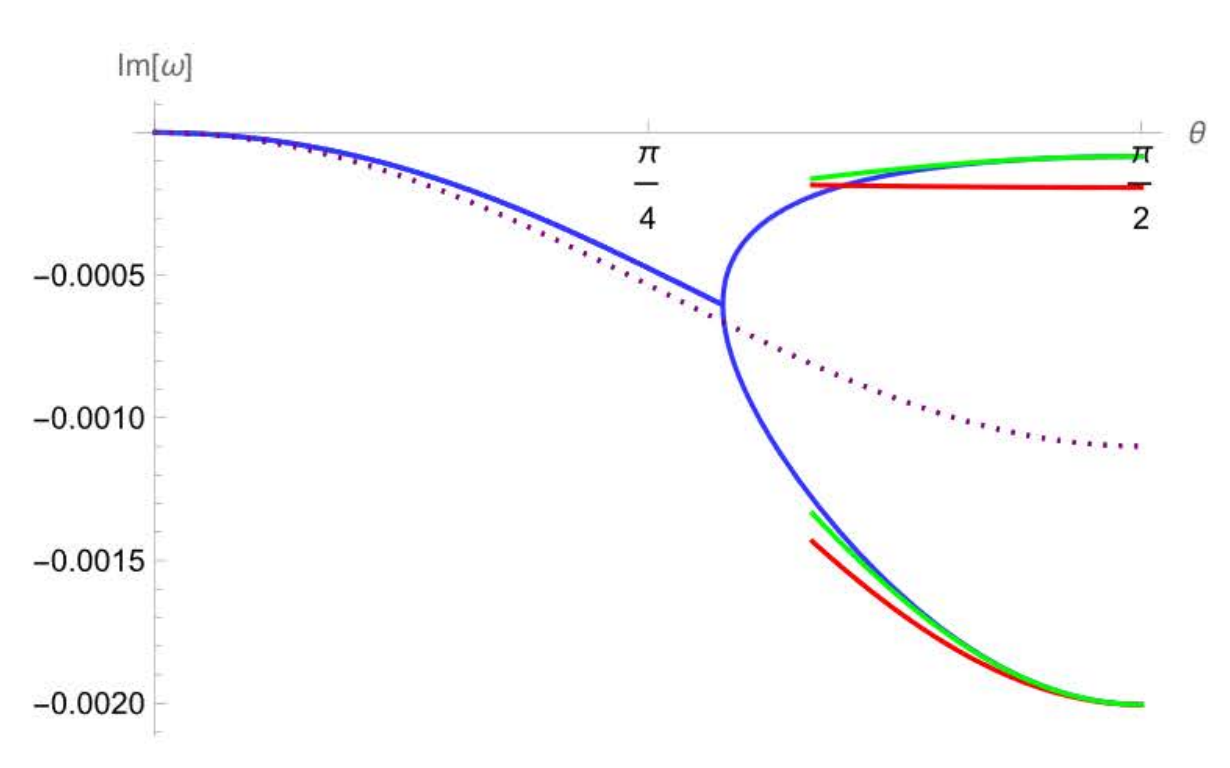}
    \caption{The dispersion relation for the slow magnetosonic modes. Blue curves show the ``exact solution'' obtained by \textit{Mathematica}.  
    Red-solid and dotted curves show the expansions with respect to $\cos\theta$ in Eq.~(\ref{eq:sonic-small-cos}) and the small $k$ in Eq.~(\ref{eq:ws}), respectively. 
    The deviation at $\theta = \pi/2$ is only due to the terms beyond the $k^2$ order. 
    } 
    \label{fig:Sonic}
\end{figure}
}

\begin{figure}
    \centering
    \includegraphics[width=\hsize]{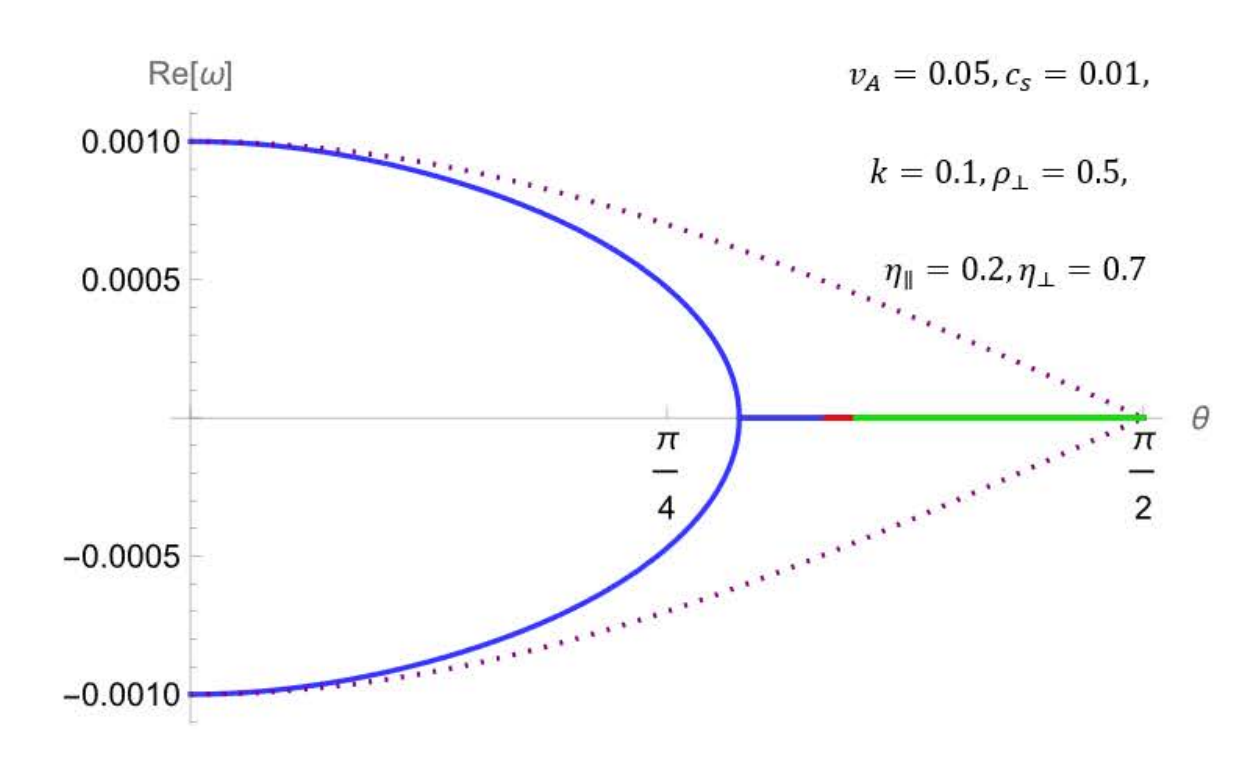}
    \\
 \includegraphics[width=\hsize]{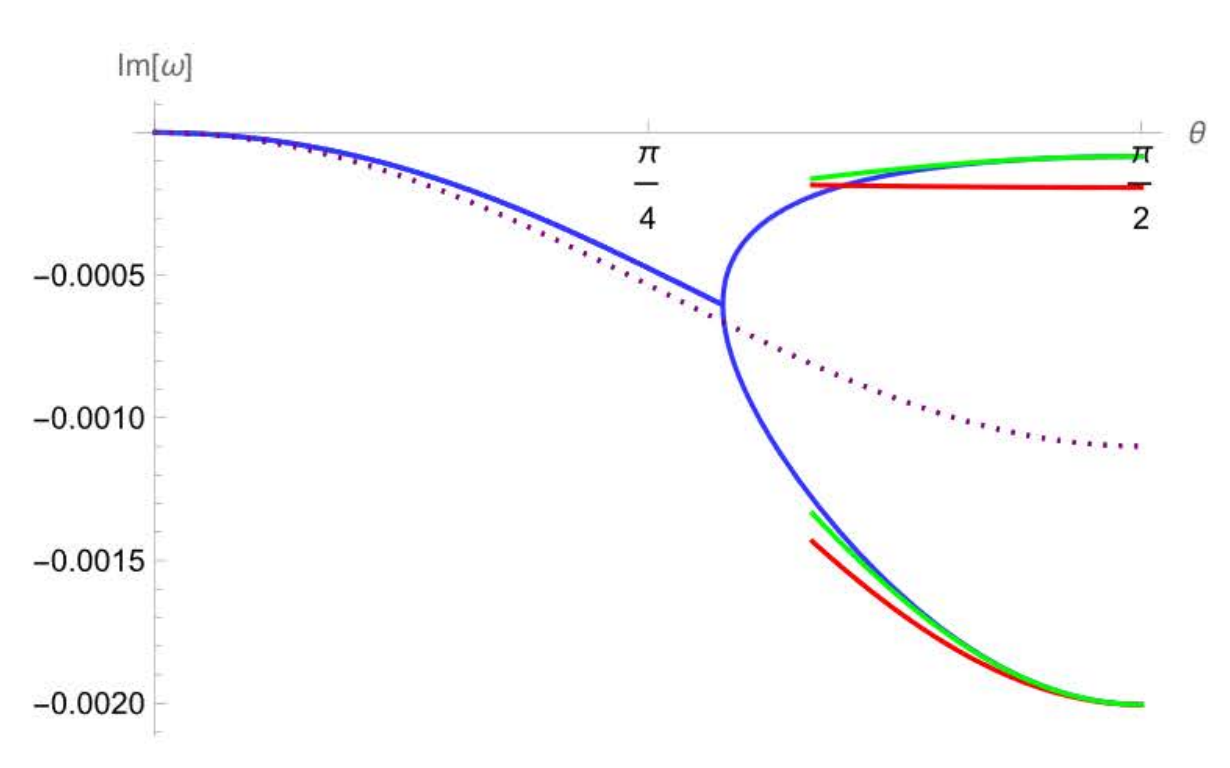}
    \caption{The dispersion relation for the slow magnetosonic modes. Blue curves show the ``exact solution'' obtained by \textit{Mathematica}. 
    Dotted curves show the small-$k$ expansion in Eq.~(\ref{eq:ws}). 
    Red and green curves show the small-cosine expansion (\ref{eq:sonic-small-cos}), instead. In the red curve, we further expand the series coefficients up to $k^2$. 
    } 
     \label{fig:Sonic}
\end{figure}

In Fig.~\ref{fig:Sonic}, we show the dispersion relations for the slow magnetosonic modes. 
The ``exact solution'' for Eq.~(\ref{eq:sonic-modes}) is shown by blue-solid curves, of which the analytic forms are obtained by an automated command in \textit{Mathematica}; 
The complicated expressions are not useful to be shown here. 
Note that the ``exact solution'' contains the higher-order corrections beyond the $k^2$ order and does not mean that it is the true goal for the first-order hydrodynamics. 
As the angle approaches the right angle $\theta = \pi/2$, the small $k$ expansion, shown by dotted curves, deviates from the ``exact solution'' even at a fixed small value of $k$. 
Instead of the small $k$ expansion, the small-cosine expansion (\ref{eq:sonic-small-cos}) should be effective in this region as discussed above. 
The green curves show the small-cosine expansion (\ref{eq:sonic-small-cos}) without an expansion for the momentum $k$, while the red curves show the same expansion but with a further expansion of the series coefficients up to $k^2$. 
Both curves well reproduce the branching in the imaginary part near the right angle. 
At $\theta = \pi/2$, i.e., $\cos\theta=0$, the deviation between the green and red curves solely comes from the choice of $\tilde \omega_i$ with or without the higher-order terms in $k$ in 
Eq.~(\ref{eq:sonic-small-cos}). 
When $\cos\theta \not=0$, the coefficients for the $\cos^2\theta$ corrections also contain the $k$ dependence in $X_i$ as well as $\tilde \omega_i$ in both the numerator and the denominator.

\cout{
On the other hand, we have confirmed that, if the ``exact solutions'' for Eq.~(\ref{equation_pi/2}) at $\theta=\pi/2$ are inserted into 
$\tilde \omega_i$, 
one finds a precise agreement between the blue-solid and dotted curves. 
However, this agreement is a result of the corrections beyond the $k^2$ order, and does not mean an improvement in accuracy for the first-order hydrodynamics. 
Therefore, inserting the solutions ($\ref{eq:sonic-disp-limit}$) suffices for the first-order hydrodynamics. 
}

\cout{
One can notice a similar issue of expansion in Alfv\'{e}n modes also appears in the slow magnetosonic wave from Fig.\ref{fig:Sonic}. When $\theta$ is smaller than $\frac{\pi}{4}$, the expansion of $k$ agrees well with exact solution. But as $\theta$ approach $\frac{\pi}{2}$, the small k expansion deviates from the curve; instead, the expansion of $\cos\theta$ is effective in this region. As a conclusion, for slow modes the expansion of k is effective only when the included angle between wave vector and magnetic field is not very large.

}

Next, we comment on extracting magnitudes of the transport coefficients by comparing the linear-mode solutions with experiments/observations. 
Remarkably, the general solution obtained in Eq.~(\ref{eq:ws}) is necessary to determine the cross bulk viscosity $\zeta_\times$, which cannot be determined with the limiting solutions at $\theta = 0, \pi/2$ \cite{Grozdanov:2016tdf}. 
It is a cross quantity between the parallel and perpendicular directions with respect to the magnetic field, and cannot be induced along a single direction at $\theta \not = 0$ or $\pi/2$ [see Eqs.~(\ref{eq:sonic-damping-perp}) and (\ref{eq:sonic-damping-para}) below]. 
Yet, even with the complete solution, one cannot determine a separation between 
$\eta'_\perp$ and $\zeta'_\perp$ that appears only in the sum in Eq.~(\ref{eq:M1}) and the general solutions (\ref{eq:W-sol}) accordingly. 
In the Alfv\'{e}n modes (\ref{eq:sol-Alfven}), 
$\eta'_\perp$ and $\rho'_\para$ also appear in the sum in the small $k$ expansion. 
However, the dependence on these two transport coefficients is split in Eq.~(\ref{eq:sol-Alfven-limit}) at $\theta = \pi/2$, and they can be determined with the linear-mode solutions. 
The other pair $\eta'_\perp$ and $\rho'_\para$ appear separately in the magnetosonic modes in Eq.~(\ref{eq:sonic-disp-limit}). 



Before closing this section, 
it is also instructive to confirm the limit of $\theta \to 0, \pi$, i.e., $k_\perp \to 0$. As mentioned below Eq.~(\ref{eq:sonic-velocities}), one of the pairs should become degenerate with the Alfv\'{e}n modes because of the rotational symmetry. 
When $ \theta = 0 $, we have 
\begin{subequations}
\label{eq:sonic-damping-para}
\begin{eqnarray} 
W_1(\theta =0)   \= \eta'_\para   c_s^2  + \rho'_\perp c_s^2   
+  \zeta'_\para   \frac{v_A^2}{1-v_A^2}  
,  
\\
W_2(\theta =0) \=  \eta'_\para  + \rho'_\perp  
+  \frac{\zeta'_\para}{1-v_A^2}  
. 
\end{eqnarray}
\end{subequations}
Then, we find that, when $ c_s \geq v_A$, 
\begin{eqnarray}  
w_1 (\theta =0) =  \frac{\zeta'_\para }{2(1-v_A^2)}  
, \,
w_2 (\theta =0) = \frac12 ( \eta'_\para + \rho'_\perp   )
\end{eqnarray} 
and that, when $ c_s < v_A$, 
\begin{eqnarray}  
w_1(\theta =0) = \frac12 ( \eta'_\para + \rho'_\perp   )
, \quad 
w_2 (\theta =0)= \frac{\zeta'_\para }{2(1-v_A^2)}  
.
\end{eqnarray} 
In both cases, either of the pairs becomes degenerate with the Alfv\'{e}n modes (\ref{eq:sol-Alfven}). 
The other pair is the sound modes damped by the bulk viscosity. 
The shear viscosity does not contribute to the damping rate in this limit, different from the usual sound modes in the absence of a magnetic field.

\section{Causality and stability}

\label{sec:s-and-c}

In the last section, we have obtained the pairs of the Alfv\'{e}n waves (\ref{eq:sol-Alfven}) and the slow and fast magnetosonic waves (\ref{eq:sonic-solutions}). 
In this section, we show that the phase velocities of the Alfv\'{e}n and magnetosonic waves are always smaller than the speed of light and that the first-order derivative corrections in those solutions (\ref{eq:sol-Alfven}) and (\ref{eq:sonic-solutions}) always act as damping factors as long as the transport coefficients satisfy the inequalities (\ref{inequality}) required by the second law of thermodynamics. 
The former implies causality in the ideal MHD. 
The latter implies a stability of equilibrium state in the {\it fluid rest frame} where the linear-mode analysis has been performed in the last section.

However, the above two properties in general do not guarantee causality beyond the ideal order or stability in an arbitrary Lorentz frame.  
We briefly discuss causality and stability of relativistic MHD along with the recent developments in the literature.

\subsection{The Alfv\'{e}n and magnetosonic velocities in the ideal MHD}

First, we focus on the linear terms in $k$, putting the higher-order terms aside. This is the ideal MHD limit. 
We assume that the sound velocity satisfies an inequality 
\begin{eqnarray}
\label{eq:ineq-sound}
0 \leq c_s \leq 1 
.
\end{eqnarray} 
We also assume that the Alfv\'{e}n velocity (\ref{eq:Alfven-velocity}) satisfies an inequality 
\begin{eqnarray}
\label{eq:ineq-Alfven}
  0  \leq v_A \leq 1 .
\end{eqnarray}
The latter inequality is evident when the energy density and pressure are separated as in Eq.~(\ref{relations}), where the Alfv\'{e}n velocity reads $v_A^2 = \frac{B^2/\mu_m}{ \epsilon+P+B^2/\mu_m} $. 
This inequality should hold unless a strong coupling between the matter and magnetic components significantly reduces the total energy density and pressure.

Under the above inequalities, one can show that the velocities of the magnetosonic waves (\ref{eq:sonic-velocities}) satisfy the inequalities   
\begin{eqnarray}
\label{eq:ineq-sonic}
0 \leq v_2 \leq c_s \leq v_1 \leq 1  \  {\rm and} \ 
0 \leq v_2 \leq v_A \leq v_1  \leq 1 .
\end{eqnarray}
Here, the relative magnitude between $ c_s$ and $ v_A$ is not assumed. 
A straightforward proof is given in Appendix~\ref{app:causality}. 
The inequalities in Eqs.~(\ref{eq:ineq-Alfven}) and (\ref{eq:ineq-sonic}) indicate that the Alfv\'{e}n and magnetosonic waves propagate with subluminal speeds in the ideal MHD. 
{The inequality (\ref{eq:ineq-sonic}) is consistent with Proposition 2.4 in Ref.~\cite{Anile_1990}.}

\cout{

One can show that 
\begin{eqnarray}
\label{eq:vel-causal}
v_{1,2} \leq 1 \, .
\end{eqnarray}
This can be done by examining if $\V^2 \pm \sqrt{ \V^4 - 4 v_A^2 c_s^2 \cos^2 \theta} \leq 2 $. 
Comparing the two sides, one find that 
\begin{eqnarray}
(2- \V^2)^2 - \sqrt{ \V^4 - 4 v_A^2 c_s^2 \cos^2 \theta} ^2 
\= -4 ( 1- v_A^2) ( 1- c_s^2 )
\nnb
&\leq& 0
\end{eqnarray}
for $ v_A^2 , \ c_s^2 \leq 1 $. 
This means Eq.~(\ref{eq:vel-causal}). 

}

\subsection{Dissipative nature of the first-order corrections in the fluid rest frame}

Next, we focus on the $k^2$ terms in the dispersion relations. 
Since all the present transport coefficients are dissipative in nature, the Alfv\'{e}n and magnetosonic waves are expected to acquire damping effects. 
We show that the pure imaginary $k^2$ terms, which have been found in the previous section, take definite signs as long as the transport coefficients satisfy the inequalities (\ref{inequality}) required by the second law of thermodynamics.

In the Alfv\'{e}n waves (\ref{eq:sol-Alfven}), it is clear that the pure imaginary coefficient in front of $k^2$ is always negative for $\eta_{\para,\perp} \geq 0$ and $\rho_{\para,\perp} \geq 0$ required by the inequalities (\ref{inequality}).

The magnetosonic waves also have the pure imaginary corrections at the $k^2$ order. 
The explicit forms of $w_{1,2}$ in Eq.~(\ref{eq:ws}) are given as 
\begin{widetext}
\begin{eqnarray}
    w_{1,2} (\theta)
\=  \frac{\mp 1}{2(v_1^2 -v_2^2)} \Big[ \, \eta'_\para \Big( c_s^2  \cos^2 (2\theta)   - v_{1,2}^2
+  \frac{(1- v_{1,2}^2 )v_A^2}{1-v_A^2} \sin^2 \theta    \Big)
+ \rho'_\perp  \Big( c_s^2  - v_{1,2}^2
+ \frac{ (1 - v_{1,2}^2 ) v_A^2 c_s^2}{1-v_A^2}  \sin^2 \theta \Big)
\nnb
&&
+ ( \eta'_\perp + \zeta'_\perp )
( c_s^2 \cos^2 \theta - v_{1,2}^2 ) \sin^2 \theta  
+ \zeta'_\para \frac{ \V^2 - v_{1,2}^2 - c_s^2 \cos^2 \theta  }
{1-v_A^2} \cos^2 \theta   
 - 2 \zeta'_\times c_s^2   \cos^2 \theta  \sin^2 \theta  
 \, \Big] 
 \label{eq:w}
 ,
\end{eqnarray}   
\end{widetext}
where the upper and lower signs are for $w_1$ and $w_2$, respectively. 
As detailed in Appendix~\ref{app:stability}, 
one can show that 
\begin{eqnarray}
\label{eq:w>0}
    w_{1,2} (\theta) \geq 0 ,
\end{eqnarray}
for any angle $\theta$. 
This means that the pure imaginary corrections always take negative signs [see the conventions in Eq.~(\ref{eq:sonic-solutions}).] 
We note that the semipositivity of $ w_{1,2}$ can be shown {\it irrespective of} the sign of $\zeta'_\times$ as long as the transport coefficients satisfy the inequalities (\ref{inequality}), none of which indeed specifies the sign of $\zeta'_\times$.

When the angle approaches $\theta = \pi/2$, one should refer to the small-cosine expansions in Eqs.~(\ref{eq:Alfven-various-expansion}) and (\ref{eq:sonic-disp-limit}) or (\ref{eq:sonic-small-cos}). 
The leading-order terms of order $\cos^0\theta$, i.e., when $\theta=\pi/2$, indicate purely diffusive modes with definite signs. 
The cosine corrections are expected to be smaller than these leading-order terms within the regions of validity for the cosine expansions. 
Then, the signs should remain definite as seen in Figs.~\ref{fig:Alfven} and \ref{fig:Sonic} with the corrections.

The above inequalities imply that both the Alfv\'{e}n and magnetosonic waves are damped out by exponential factors for an observer in the fluid rest frame where the linear-mode analysis has been performed. 
However, those inequalities are not sufficient conditions for stability in an arbitrary Lorentz frame, but are necessary conditions. 
We discuss stability and causality conditions in a more general perspective below. 


\cout{
\begin{subequations}
\label{eq:w}
    \begin{eqnarray}
        \label{eq:w1}
w_1 
\=  \frac{-1}{2(v_1^2 -v_2^2)} \Big[ \, \eta'_\para \Big( c_s^2  \cos^2 (2\theta)   - v_1^2
+  \frac{(1- v_1^2 )v_A^2}{1-v_A^2} \sin^2 \theta    \Big)
+ \rho'_\perp  \Big( c_s^2  - v_1^2
+ \frac{ (1 - v_1^2 ) v_A^2 c_s^2}{1-v_A^2}  \sin^2 \theta \Big)
\nnb
&&
+ ( \eta'_\perp + \zeta'_\perp )
( c_s^2 \cos^2 \theta - v_1^2 ) \sin^2 \theta    
+ \zeta'_\para \frac{ \V^2 - v_1^2 - c_s^2 \cos^2 \theta  }
{1-v_A^2} \cos^2 \theta   
 - 2 \zeta'_\times c_s^2   \cos^2 \theta  \sin^2 \theta  \, \Big]
 \\
\label{eq:w2}
w_2
\= \frac{+1}{2(v_1^2 -v_2^2)} \Big[ \, \eta'_\para \Big( c_s^2  \cos^2 (2\theta)  - v_2^2 
+  \frac{(1- v_2^2 )v_A^2}{1-v_A^2} \sin^2 \theta  \Big)
+ \rho'_\perp  \Big( c_s^2   - v_2^2 
+ \frac{ (1 - v_2^2 ) v_A^2 c_s^2}{1-v_A^2}  \sin^2 \theta  
\Big)
\nnb
&&
+ ( \eta'_\perp + \zeta'_\perp) ( c_s^2 \cos^2 \theta - v_2^2 ) \sin^2 \theta   
+ \zeta'_\para \frac{ \V^2 - v_2^2- c_s^2 \cos^2\theta }{1-v_A^2} \cos^2 \theta  
- 2 \zeta'_\times c_s^2 \cos^2 \theta  \sin^2 \theta  \, \Big]
    \end{eqnarray}
\end{subequations}

}

\cout{

The terms associated with the bulk viscosities have the form 
\begin{eqnarray}
w_\zeta = s \zeta_\para +  t \zeta_\perp +  q \zeta_\times 
\, .
\end{eqnarray}
When $ s, t \geq 0 $, we have 
\begin{eqnarray}
w_\zeta \geq 2 \sqrt{ s t \zeta_\para  \zeta_\perp} 
+  q \zeta_\times 
\, .
\end{eqnarray}
Comparing the two terms, we have 
\begin{eqnarray}
4 s t \zeta_\para  \zeta_\perp - ( q \zeta_\times )^2
\geq (4 st - q^2) \zeta_\times ^2 
= 0
\, ,
\end{eqnarray}
where we used the inequality from the thermodynamic constraint 
and the explicit forms of $w_{1,2} $ that, 
in both cases, lead to 
\begin{eqnarray}
4 st - q^2  = 0
\, .
\end{eqnarray}
Therefore, one can conclude that $ w_\zeta \geq 0$ 
as long as $ s, t \geq 0 $. 


}

\subsection{Covariant stability in anisotropic systems}

\label{sec:causal-stability}

It has been known that diffusive modes in the first-order hydrodynamics are acausal and that such diffusive modes, which are damped out in the fluid rest frame, can be transformed into growing modes in a general Lorentz frame \cite{Hiscock:1985zz}. 
Relativistic hydrodynamic theories containing such instability may not work in practice
because the stability of local equilibria in a certain reference frame, e.g., a lab frame as often interested, is not guaranteed. 
Therefore, it is important to understand the origin of the instability and the necessary and/or sufficient conditions for the {\it covariant stability} where the local equilibria are stable in {\it any} Lorentz frame.

Here, we briefly discuss the covariant stability for MHD with a slight extension of the recent discussions by Gavassino for isotropic systems \cite{Gavassino:2021owo, Gavassino:2023myj}. 
We write the solutions for the linear perturbations (\ref{eq:linear-solutions}) in a covariant form 
\begin{eqnarray}
\label{eq:sol-general}
\Psi = \tilde \Psi e^{-i k_\mu x^\mu}
,
\end{eqnarray}
where $\Psi$ is an eigenvector 
and the corresponding momentum $k^\mu 
= (\omega, \bk)$ satisfies a dispersion relation obtained from the (linearized) hydrodynamic equations. 
However, as we have seen in the previous sections, the dispersion relations are not given in Lorentz-invariant forms such as the free-particle on-shell conditions in quantum field theories, 
which is due to the derivative expansion. 
Therefore, we need to understand how the dispersion relations are transformed by a Lorentz boost.

First, notice that, in dissipative hydrodynamics, the spatial component of the momentum $\bk$ develops an imaginary part in a general Lorentz frame  due to the mixing between $\omega$ and $\bk$ under Lorentz boosts. 
For the general solution (\ref{eq:sol-general}) to be covariantly stable, 
one should have $\Im m \, [k_\mu x^\mu] \leq 0$. 
For this condition to be satisfied for observers in the forward light cone, the imaginary part of $k^\mu$ should be a vector lying outside the forward light cone, requiring that 
\begin{eqnarray}
&&
\Im m \, \omega \leq - |\Im m \, \bk|
\quad {\rm or} \quad 
\\
&&
(\Im m \, k^\mu ) (\Im m \, k_\mu )
= - (\Im m \, \omega )^2 
+ |\Im m \, \bk|^2
\geq 0 \nn
.
\end{eqnarray}
Combining the above inequalities, one finds a necessary condition for the covariant stability 
\begin{eqnarray}
\label{eq:omega-k-anisotropic}
\Im m \, \omega \leq |\Im m \, \bk|
\, .
\end{eqnarray}
This is an extension of the condition for an isotropic system obtained by Heller \textit{et al}. from the analytic property of a general retarded propagator in causal quantum field theory theories \cite{{Heller:2022ejw}} and interpreted by Gavassino as the covariantly stability condition for relativistic dissipative hydrodynamics \cite{Gavassino:2023myj}. 
The inequality (\ref{eq:omega-k-anisotropic}) is a stronger condition than that for isotropic systems where there is essentially only one independent component of $\bk$. 
{Namely, while it was a relation between the frequency $\omega(k,0,0)$ and one component of the momentum $k$ for dispersion relations in isotropic systems, dispersion relations in anisotropic systems are relations between 
the frequency $\omega(k_x,k_y,k_z)$ and independent momentum components, and so is the condition (\ref{eq:omega-k-anisotropic}). 
}
For MHD, two of the three components should be treated independently as there remains a rotational symmetry around the magnetic-field direction.  

When there are dissipative effects, i.e., $\Im m \, \omega \not = 0$, with $\Im m \, \bk =0$ in the fluid rest frame, one may consider successive multiple Lorentz boosts: The first boost generates an imaginary part of $\bk$ due to the mixing with $\omega$, and the subsequent boosts require the inequality (\ref{eq:omega-k-anisotropic}) for the causal stability. 
The result of these successive boosts is not equivalent to that of a single boost by a sum of the boost velocities, because of the non-Abelian nature of the Lorentz group. 
In analogy with the discussion about the Thomas precession \cite{jackson1999classical}, such a sequence of Lorentz boosts is required to move from one Lorentz frame to another, e.g., from the rest frame of a fluid cell to the lab frame, {\it when the fluid cell is accelerated}.

It is instructive to explicitly see the occurrence of instability when the inequality (\ref{eq:omega-k-anisotropic}) 
is {\it not} satisfied \cite{Gavassino:2023myj}. 
In fact, the dispersion relations turn into unstable ones when boosted by a velocity 
\begin{eqnarray}
\bv = \frac{ \Im m \, 
\bk}{\Im m \, \omega}  
\, ,
\end{eqnarray}
that satisfies $|\bv| \leq 1$ when the inequality (\ref{eq:omega-k-anisotropic}) 
is not satisfied.  
Boosting the imaginary part of the momentum, one finds that 
$\Im m \, k^\mu \to 
\Im m \, k^{\prime\mu} $, where 
\begin{eqnarray}
\Im m \, k^{\prime\mu} \= 
\gam (\Im m \, \omega - \bv\cdot \Im m \, \bk,
\nnb
&&
-  \gam \bv \Im m \, \omega + \Im m \, \bk 
+ \frac{(\gam-1)  \bv \cdot \Im m \, \bk }{v^2} \bv  ) 
\nnb
\= 
\gam ( \frac{ (\Im m \, \omega)^2 
- | \Im m \, \bk |^2 
}{\Im m \, \omega}  ,
{\bm 0} ) 
\, .
\end{eqnarray}
This means that an observer in the new frame claims 
the existence of an unstable Fourier mode because 
$\Im m \, \omega' > 0 $ and $ \Im m \, \bk ' =0$.  
Therefore, the inequality (\ref{eq:omega-k-anisotropic}) is indeed necessary for the covariant stability.

If the inequality (\ref{eq:omega-k-anisotropic}) is satisfied in one of the Lorentz frames, it is, by construction, satisfied in all the Lorentz frames connected by Lorentz boosts. 
Then, the covariant stability will be fulfilled {\it if} signals only reach observers inside the forward light cone, i.e., if theories respect causality (see a theorem in Sec.~III.B in Ref.~\cite{Gavassino:2021kjm}). 
However, it has been known that, in dissipative hydrodynamics, signals reach observers outside the forward light cone. 
The acausal tails of dissipative modes are not only illegitimate in relativity in the first place but also observed as unstable modes outside the light cone (see a theorem in Sec.~III.A in Ref.~\cite{Gavassino:2021kjm}). 
This is because, for a spacelike separation $(y-x)^2 >0$, there is a Lorentz boost that inverts the chronicle ordering as $y^0-x^0 \to y^{\prime\, 0}-x^{\prime\, 0} = -(y^0-x^0)$, making the meaning of dissipation and growth observer-dependent concepts. 
Therefore, the covariant stability is fulfilled only in causal theories. 
The first-order MHD is stable in the fluid rest frame as shown above, but is not causal. 
The subluminal magnetosonic velocities (\ref{eq:ineq-sonic}) only serve as necessary conditions for causality once the dissipative effects are included.

Causality is often a consequence of subtle cancellation among the acausal tails leaking across the light cone. 
A well-known example is the Klein-Gordon field in quantum field theory \cite{Peskin:1995ev}. 
In Ref.~\cite{Gavassino:2023mad}, it is stated that any dispersion relation can leak across the light cone unless a medium is not dispersive or, in other words, dispersion relations are polynomials of the first order at most, i.e., $\omega(k) = a_0  + a_1 k$ (see also Ref.~\cite{Heller:2022ejw}). This implies that inspecting each dispersion relation alone does not guarantee causality.\footnote{The Israel-Stewart theory is one of such cases where the dissipative corrections, e.g., the viscous tensors, are promoted to independent variables, and the conservation laws are cast into a larger set of the first-order differential equations in {\it both} space and time (if the vorticity terms are neglected) \cite{Bemfica:2020xym}. 
Otherwise, one can convert a set of second-order differential equations to that of first-order differential equations by introducing auxiliary fields \cite{Bemfica:2020zjp}.} 
Notions of velocities for a single dispersion relation, such as the front velocity, the group velocity, and the phase velocity, may be useful for screening apparently acausal theories, but do not serve as a sufficient causality test; Besides, it should be noticed that the font velocity is defined at the ultraviolet limit $k\to \infty$ outside the hydrodynamic regime. 
Instead, it will be useful to investigate causal structures of a set of partial differential equations with the method of characteristics (see, e.g., Refs.~\cite{Izumi:2014loa, Bemfica:2017wps} and references therein), though it will require more efforts in future works. 
It is worth adding that the covariantly stable condition (\ref{eq:omega-k-anisotropic}) is derived independently of any notion of velocities (see also Theorem 2 in Ref.~\cite{Gavassino:2023myj} where a criterion of causality is manifestly implemented without any notion of velocities).

\cout{
\begin{eqnarray}
\tilde G_R (k) = \int d^4 x G_R(x^\mu) e^{-i k_\mu x^\mu}
\end{eqnarray}
with the mostly plus metric convention. 
For the retarded correlator $\tilde G_R (k)$ to be bounded in the momentum space, 
the regarded correlator $G_R(x^\mu)$ is bounded in any Lorentz frame and the imaginary part of the momentum vector should lay outside the forward light cone.  
For the retarded correlator to be stable in the momentum space, the 
}

\section{Conclusion and outlook}\label{sec:conclusion}

In this paper, we investigated linear waves in relativistic magnetohydrodynamics in detail. 
Especially, in Sec.~\ref{sec:magneto-sonic}, we provided a simple and general analytic algorithm for the solution search. 
Based on this algorithm, we showed analytic solutions for the magnetosonic waves that have been missing in the literature for a long time. 
The algorithm can be applied to other hydrodynamic equations or any general set of equations based on a derivative expansion. 
We will provide an application elsewhere \cite{Fang:2024sym}. 
Also, while we focused on the Landau frame in the present work, it is interesting to investigate analytic solutions in a general choice of hydrodynamic variables (cf. Ref.~\cite{Armas:2022wvb}). 

On the other hand, we also found that the small-momentum expansion for the solutions breaks down in MHD when the momentum direction is nearly or exactly perpendicular to an equilibrium magnetic field. 
This issue occurs in both the Alfv\'{e}n and magnetosonic waves and stems from the competition between two small quantities involved in the solutions that are the momentum and the trigonometric functions representing the spatial anisotropy in MHD, i.e., a cosine function in the present convention. 
When the cosine becomes small near the right angle, we found that the higher-order terms in the small-momentum expansion diverge, spoiling the small-momentum expansion. 
The breakdown of the small-momentum expansion can be a general issue emerging in anisotropic systems. 
We provided alternative expressions of the solutions based on the small-cosine expansion in Eqs.~(\ref{eq:Alfven-various-expansion}) and (\ref{eq:sonic-small-cos}) that work accurately near the right angle as shown in Figs.~\ref{fig:Alfven} and \ref{fig:Sonic}.

Last, we investigated the issues of  causality and stability in the first-order relativistic MHD based on the analytic solutions. 
We showed that the Alfv\'{e}n and magnetosonic velocities are less than the speed of light and that the first-order corrections always act as damping effects in the fluid rest frame. 
As mentioned in Sec.~\ref{sec:causal-stability}, these conditions are, however, not sufficient for the covariant stability, i.e., the stability in all the Lorentz frames. 
The main and general reason is that dissipative hydrodynamics exhibits acausal propagation across the forward light cone {(even though phase velocities appearing in the ideal order are subluminal)}. 
Such acausal signals can be observed as unstable modes in the spacelike regions, indicating an intimate connection between the issues of causality and stability. 
So far, the method of moment expansion, which leads to the Israel-Stewart theory, has been invoked in Refs.~\cite{Denicol:2018rbw, Denicol:2019iyh} to formulate causal and stable MHD (see also Refs.~\cite{Hattori:2022hyo} for a review). 
{
Linear stability analyses in the fluid rest frame were performed for the Israel-Stewart extension \cite{Biswas:2020rps} and for a general matching condition \cite{Armas:2022wvb}. 
One of the open issues in these stability analyses is solution searches at a general angle other than $\theta=0$ and $\pi/2$. 
Applying our method, one could obtain stronger constraints on transport coefficients for stability and causality discussed in Refs.~\cite{Biswas:2020rps, Armas:2022wvb}. 
The breakdown of the momentum expansion discussed in our paper may also affect the stability and causality conditions. 
}
It is yet left as an open question to formulate covariantly stable MHD based on the magnetic-flux conservation (cf. Sec.~\ref{s2}). 
Other future works include computation of the transport coefficients (see \cite{Hattori:2016cnt, Hattori:2016lqx, Hattori:2017qih, Li:2017tgi, Fukushima:2017lvb, Kurian:2018dbn, Li:2018ufq, Fukushima:2019ugr, Astrakhantsev:2019zkr, Fukushima:2021got, Peng:2023rjj} for recent studies). 
These developments will promote further numerical studies \cite{Inghirami:2016iru, Inghirami:2019mkc, Nakamura:2022idq, Nakamura:2022wqr, Nakamura:2022ssn}.

\acknowledgments
We thank Yi-hui Tu, Shi Pu, and Dong-Lin Wang 
for useful discussions. 
This work is partially supported by the JSPS  KAKENHI under Grants No. 20K03948 and No. 22H01216, and the start-up Grant No. XRC-23112 of Fuzhou University.

\appendix

\section{INEQUALITIES}

\subsection{Inequalities for the Alfv\'{e}n and magnetosonic velocities}

\label{app:causality}

We assume that the equations of state satisfy 
the inequalities $0 \leq c_s \leq 1 $ and $ 0 \leq v_A \leq 1 $ as stated in Eqs.~(\ref{eq:ineq-sound}) and (\ref{eq:ineq-Alfven}). 
Then, we show that the velocities of the magnetosonic waves (\ref{eq:sonic-velocities}) satisfy the inequalities (\ref{eq:ineq-sonic}), i.e.,  
\begin{eqnarray}
\label{eq:ineq-velocity}
0 \leq v_2 \leq c_s \leq v_1 \leq 1  \quad {\rm and} \quad 
0 \leq v_2 \leq v_A \leq v_1  \leq 1 .
\nnb
\end{eqnarray}
Here, the relative magnitude between $ c_s$ and $ v_A$ is not assumed.

{

First of all, one can show that $v_{1,2}$ are real-valued quantities. 
This amounts to showing positivity in the most inside of the square root: 
\begin{eqnarray}
&&
    \mathcal V^2-4v_A^2 c_s^2 \cos^2\theta
\nnb
&\geq& (c_s^2+v_A^2)^2+c_s^4 v_A^4 \sin^4\theta-4c_s^2 v_A^2 \sin^2\theta-4 v_A^2 c_s^2 \cos^2\theta
\nonumber\\
&=&(c_s^2-v_A^2)^2+c_s^4 v_A^4 \sin^4\theta
\nonumber\\
&\geq& 0 \, ,
\end{eqnarray}
where we used $c_s^2, \, v_A^2 \leq 1$. 
}

It is easy to show that $\V^2 \geq 0 $ 
by comparing the magnitudes of the two terms in $\V^2 $ as 
\begin{eqnarray}
&&
(c_s^2 + v_A^2)^2 - (c_s^2 v_A^2 \sin^2 \theta)^2 
\nnb
\= c_s^4 + v_A^4 + c_s^2  v_A^2 ( 2 - c_s^2 v_A^2 \sin^4 \theta)
\geq 0
. 
\end{eqnarray}
Then, it is obvious that $v_{1,2} \geq 0 $. 

Next, we show that $v_{1,2} \leq 1 $. 
Comparing the two sides, one finds that 
\begin{eqnarray}
(2-\V^2)^2 - \sqrt{\V^4 - 4 v_A^2 c_s^2 \cos^2 \theta} ^{\, 2}
\= -4 ( 1- v_A^2) ( 1- c_s^2 )
\nnb
&\leq& 0
.
\end{eqnarray}
Note also that $ 2-\V^2 = ( 2 -c_s^2 -v_A^2) + c_s^2 v_A^2 \sin^2\theta \geq 0 $. 
Then, one can conclude that $v_{1,2} \leq 1 $.

Lastly, we show the relative magnitudes of 
$v_{1,2} $ to $c_s$ and $v_A $. 
The difference between $v_{1,2} $ and $c_s$ reads 
\begin{eqnarray}
\label{eq:c-v}
c_s^2 - v_{1,2}^2 \= ( c_s^2 - \frac12 \V^2) 
\mp \frac{1}{2}  \sqrt{ \V^4 - 4 v_A^2 c_s^2 \cos^2 \theta } 
. 
\end{eqnarray}
The relative magnitudes of the two terms is examined as 
\begin{eqnarray}
&&
 \frac{1}{2^2}  \big( \V^4 - 4 v_A^2 c_s^2 \cos^2 \theta \big)
 - ( c_s^2 - \frac12 \V^2)^2 
 \nnb
\= ( - v_A^2  \cos^2 \theta - c_s^2 +  \V^2 ) c_s^2
\nnb
\= (  1- c_s^2  ) v_A^2 c_s^2 \sin^2 \theta  
\nnb
& \geq & 0 .
\end{eqnarray}
Therefore, the sign of the difference in Eq.~(\ref{eq:c-v}) 
is determined by that of the square-root term 
regardless of the sign of the other term. 
Then, one can conclude that $  v_2 \leq c_s \leq v_1 $. 
By the same token, one can examine the difference 
\begin{eqnarray}
v_A^2 - v_{1,2}^2 \= ( v_A^2 - \frac12 \V^2) 
\mp \frac{1}{2}  \sqrt{ \V^4 - 4 v_A^2 c_s^2 \cos^2 \theta }  
. 
\end{eqnarray}
The relative magnitude of the two terms is found to be 
\begin{eqnarray}
&&
 \frac{1}{2^2}  \big( \V^4 - 4 v_A^2 c_s^2 \cos^2 \theta \big)
 - ( v_A^2 - \frac12 \V^2 )^2 
 \nnb
\= ( - c_s^2  \cos^2 \theta - v_A^2 +  \V^2 ) v_A^2
\nnb
\= (  1- v_A^2  ) v_A^2 c_s^2 \sin^2 \theta
\nnb
& \geq & 0 .
\end{eqnarray} 
Then, one can conclude that $  v_2 \leq v_A \leq v_1 $.

Following the above proof, we conclude the inequalities (\ref{eq:ineq-sonic}).

\subsection{Dissipative corrections in the fluid rest frame}

\label{app:stability}

The next-to-leading order solutions in the magnetosonic modes (\ref{eq:sonic-solutions}) 
are obtained with $ w_{1,2}$ given in Eq.~(\ref{eq:w}). 
Here, we show that $w_{1,2} \geq 0 $ for the transport coefficients that satisfy the inequalities (\ref{inequality}) required by the second law of thermodynamics. 
In the following discussion, one can forget about the positive overall factor like
$1/(2(v_1^2 - v_2^2) )$ that is irrelevant for examining the signs of $w_{1,2}$.

We begin with the terms associated with 
the bulk viscosities $ \zeta_{\para,\perp,\times}$. 
Picking up these terms from $w_{1,2}$ in Eq.~(\ref{eq:w}), we have 
\begin{eqnarray}
w^\zeta_{1,2} := s_{1,2} \zeta_\perp
+  t_{1,2} \zeta_\para +  u  \zeta_\times 
\, ,
\end{eqnarray}
where 
\begin{eqnarray}
s_{1,2} \=  \mp ( c_s^2 \cos^2 \theta - v_{1,2}^2 ) \sin^2\theta
, \quad 
\\
t_{1,2} \= \mp
\frac{ \V^2 - v_{1,2}^2- c_s^2 \cos^2\theta }{1-v_A^2}
\cos^2 \theta 
, 
\\
u  \= \pm  2 c_s^2 \cos^2 \theta  \sin^2 \theta  .
\end{eqnarray}
One can show that $s_{1,2} \geq 0 $ 
and $t_{1,2} \geq 0 $ as we will see later. 
Assuming these positivities for the moment, one finds that 
\begin{eqnarray}
w^\zeta_{1,2} \geq 
2 \sqrt{ s_{1,2} t_{1,2} \zeta_\para  \zeta_\perp} 
+  u \zeta_\times 
\, .
\end{eqnarray}
Further examining the relative magnitude of the two terms on the right-hand side, we have 
\begin{eqnarray}
&&
4 s_{1,2} t_{1,2} \zeta_\para  \zeta_\perp 
- ( u \zeta_\times )^2
\nnb
&\geq& (4 s_{1,2}t_{1,2} - u^2) \zeta_\times ^2 
= 0
\, ,
\end{eqnarray}
where we used the inequality from the thermodynamic constraint (\ref{inequality}) and the explicit forms of $s,t,u $ that, 
in both cases, lead to $(4 s_{1,2}t_{1,2} - u^2 ) = 0 $. 
Therefore, one can conclude that $ w_{1,2}^\zeta \geq 0$ 
as long as $ s_{1,2}, t_{1,2} \geq 0 $ {\it irrespective of} the sign of $ u$.

To show that $s_{1,2} \geq 0 $, one can arrange it 
with the explicit forms of $ v_{1,2}^2$ as 
\begin{eqnarray}
&&
s_{1,2} 
= \mp \big(  c_s^2 \cos^2 \theta - \frac{1}{2}  \V^2  \big)
\nnb
&&
+ \frac{1}{2}  \sqrt{ \V^4 - 4 v_A^2 c_s^2 \cos^2 \theta }  
. 
\end{eqnarray} 
As for $t_{1,2} \geq 0 $, one can focus on the numerator 
\begin{eqnarray}
&&
(1-v_A^2) t_{1,2} 
\nnb
\= \mp( \V^2 - v_{1,2}^2- c_s^2 \cos^2\theta )
\\
\= \pm \big( c_s^2 \cos^2 \theta - \frac12 \V^2 \big)
+ \frac{1}{2}  \sqrt{ \V^4 - 4 v_A^2 c_s^2 \cos^2 \theta }  
\nn
.
\end{eqnarray}
In both cases, one can show that the square-root term is always larger than the absolute value of the first term, that is, 
\begin{eqnarray}
&&
\frac{1}{2^2} ( \V^4 - 4 v_A^2 c_s^2 \cos^2 \theta) 
- (c_s^2 \cos^2 \theta - \frac{1}{2}  \V^2 )^2
\nnb
\= -  c_s^2 \cos^2 \theta ( v_A^2 + c_s^2 \cos^2 \theta - \V^2 )
\nnb
\=  c_s^4 \sin^2 \theta \cos^2 \theta  (  1 -  v_A^2  )
\nnb
& \geq & 0 .
\end{eqnarray}
Therefore, we have shown that 
$ s_{1,2} \geq 0 $ and $t_{1,2} \geq 0$, 
and accordingly that $w_{1,2}^\zeta \geq 0 $.

We have three remaining terms associated with 
$ \eta'_\para, \rho'_\perp, \eta'_\perp$. 
The last one $\eta'_\perp $ only appears with $ \zeta'_\perp$ in Eq.~(\ref{eq:w}), 
so that we have already shown the positivity of this term just above. 
We examine the remaining two terms below. 
The coefficients in front of $\rho'_\perp $ are arranged as 
\begin{eqnarray}
\label{eq:ineq-rho} 
&&
\frac{ \mp 1 }{1-v_A^2} 
\big[ \, (1-v_A^2) (c_s^2 -v_{1,2}^2) 
\nnb
&&
+ (1 - v_{1,2}^2 ) v_A^2 c_s^2 \sin^2 \theta  \, \big]
.  
\end{eqnarray}
According to the inequalities (\ref{eq:ineq-velocity}) for the velocities, the right-hand side is 
semipositive definite for $v_2 $. 
As for $ v_1$, one can arrange the expression between the square brackets as 

\begin{eqnarray}
\label{eq:rho-ineq-1}
&& 
(1-v_A^2) (c_s^2 -v_1^2) 
- (1 - v_1^2 ) v_A^2 c_s^2 \sin^2 \theta 
\nnb
\= c_s^2 (1-v_A^2  + v_A^2  \sin ^2 \theta) - v_1^2 J
\nnb
\= \big[ c_s^2 (1-v_A^2  + v_A^2  \sin ^2 \theta) - \frac12 \V^2 J \big]
\nnb
&&
- \frac{J}{2}  \sqrt{ \V^4 - 4 v_A^2 c_s^2 \cos^2 \theta } 
,
\end{eqnarray}
where $ J = 1-v_A^2 +  v_A^2 c_s^2 \sin ^2 \theta $. 
Comparing the magnitudes of the two terms, one finds that 
\begin{eqnarray}
&&
\Big[ \, \frac{J}{2} \sqrt{ \V^4 - 4 v_A^2 c_s^2 \cos^2 \theta } \, \Big]^2
\nnb
&&
- \big[ c_s^2 (1-v_A^2  + v_A^2  \sin ^2 \theta)
- \frac12 \V^2 J \big]^2
\nnb
\=  v_A^2 c_s^2 (1-v_A^2) (1-c_s^2)^2 \sin^2 \theta 
\geq 0
\, .
\end{eqnarray}
This means that the sign of the left-hand side in Eq.~(\ref{eq:rho-ineq-1}) is determined by that of the square-root term, which is negative. 
Therefore, for both $ w_{1,2}$, one can conclude that the coefficients in front of $\rho_\perp'$ are semipositive definite in Eq.~(\ref{eq:ineq-rho}).

\cout{
\begin{eqnarray}
\label{eq:ineq-rho} 
\frac{ \mp 1 }{1-v_A^2} 
\big[ \, (1-v_A^2) (c_s^2 -v_{1,2}^2) 
+ (1 - v_{1,2}^2 ) v_A^2 c_s^2 \sin^2 \theta  \, \big]
\\
=\pm[-(1-v_A^2)c_s^2-v_A^2 c_s^2 \sin^2(\theta)] \pm(1-v_A^2+v_A^2 c_s^2 \sin^2(\theta))v_{1,2}^2
,
. 
\end{eqnarray}
We shall prove it is semipositive. And it is obvious that $(1-v_A^2+v_A^2 c_s^2 \sin^2(\theta))\geq 0$, so we shall prove:  
\begin{eqnarray}
\pm v_{1,2}^2=\frac{1}{2}(\pm \V^2+\sqrt{\V^4 - 4 v_A^2 c_s^2 \cos^2 \theta}) \geq \pm\frac{(1-v_A^2)c_s^2+v_A^2 c_s^2 \sin^2(\theta)}{1-v_A^2+v_A^2 c_s^2 \sin^2(\theta)}
.
\end{eqnarray}
Move the root to one side. Squaring each term and taking the difference, we have 
\begin{eqnarray}
 \frac{1}{2^2}  \big( \V^4 - 4 v_A^2 c_s^2 \cos^2 \theta \big)
 - \Big(\pm \frac{(1-v_A^2)c_s^2+v_A^2 c_s^2 \sin^2(\theta)}{1-v_A^2+v_A^2 c_s^2 \sin^2(\theta)}\mp\frac 12 \V^2 \Big)^2
\\
 =\frac{cs^2 (1 - cs^2)^2 v_A^2 (1 - v_A^2) \sin^2\theta}{(1-v_A^2+v_A^2 c_s^2 \sin^2(\theta))^2}\geq 0
.
\end{eqnarray}
Therefore, for both $ w_{1,2}$, one can conclude that the coefficients in front of $\rho_\perp'$ are semipositive definite. 

}

Last, the coefficients in front of $ \eta'_\para$ in Eq.~(\ref{eq:w}) can be arranged as 
\begin{eqnarray} 
&&
\frac{ \pm 1}{1-v_A^2} 
\big[ (1-v_A^2 \cos^2 \theta ) v_{1,2}^2
\nnb
&&
-  (1-v_A^2)c_s^2  \cos^2 (2\theta)  - v_A^2 \sin^2 \theta  \big] 
. 
\label{eq:eta-1}
\end{eqnarray}
Inserting the explicit forms of $v_{1,2} $, we have 
\begin{eqnarray}
&&
(1-v_A^2 \cos^2 \theta )v_{1,2}^2 - \{ (1-v_A^2)c_s^2  \cos^2 (2\theta)  + v_A^2 \sin^2 \theta\}
\nnb
&&
= K_{1,2} + L
. \nnb \label{eq:eta-2}
\end{eqnarray}
where 
\begin{subequations}
    \begin{eqnarray}
K_{1,2} \= \pm \frac{1}{2} (1-v_A^2 \cos^2 \theta )
 \sqrt{ \V^4 - 4 v_A^2 c_s^2 \cos^2 \theta }
,
\\
L \=  \frac12 (1-v_A^2 \cos^2 \theta )\V^2 
-  (1-v_A^2) c_s^2 \cos^2 (2\theta) \nnb
&-&v_A^2 \sin^2 \theta  
, \nnb
    \end{eqnarray}
\end{subequations}
where the upper and lower signs from $v_{1,2}$ are for $K_1$ and $K_2$, respectively. 
Examining the relative magnitude of the two terms, one finds that 
\begin{eqnarray} 
&&
K_{1,2}^2 - L^2
\nnb
\= \frac14 (1-v_A^2)\sin^2 (2\theta)
\big \{ v_A^2 - c_s^2(2-v_A^2) \cos (2\theta) \big\} ^2
\nnb
&\geq&  0 .
\end{eqnarray}
This inequality, together with $ (1-v_A^2 \cos^2 \theta ) \geq 0$, 
means that the overall signs in Eq.~(\ref{eq:eta-2}) are determined by 
that of $K_{1,2}$. 
Then, one can conclude that 
the coefficients in front of $ \eta'_\para$ 
are semipositive definite for both $ w_{1,2}$.

From the above, we conclude that the magnetosonic modes (\ref{eq:sonic-solutions}) 
are damped out by the semipositive damping factor $ w_{1,2}$ given in Eq.~(\ref{eq:w}). 
We emphasize that the semipositivity of $ w_{1,2}$ has been shown irrespective of the sign of $\zeta'_\times$ as long as the transport coefficients satisfy the inequalities (\ref{inequality}), none of which specifies the sign of $\zeta'_\times$.

\bibliography{ref}

\end{document}